\documentclass[11pt,a4paper]{article}
\usepackage{jheppub} 
\usepackage{bm}
\usepackage{slashed,bbold}

\newcommand{\beq}{\begin{equation}} 
\newcommand{\eeq}{\end{equation}}
\newcommand{\beqa}{\begin{eqnarray}} 
\newcommand{\eeqa}{\end{eqnarray}}

\newcommand{\A}{{\mathcal A}}
\newcommand{\cF}{{\mathcal F}}
\newcommand{\cG}{{\mathcal G}}
\newcommand{\cH}{{\mathcal H}}
\newcommand{\cP}{{\mathcal P}}
\newcommand{\cW}{{\mathcal W}}

\newcommand{\cL}{{\cal L}}

\title{Second-order partition function of a non-interacting chiral fluid in 3+1 dimensions}

\author[a]{Eugenio Meg\'{i}as,}
\author[b]{Manuel Valle}

\affiliation[a]{Grup de F\'{\i}sica Te\`orica and IFAE, Departament de F\'{\i}sica, Universitat Aut\`onoma de Barcelona, 
Bellaterra 08193 Barcelona, Spain}
\affiliation[b]{Departamento de F\'{\i}sica Te\'orica, 
Universidad del Pa\'{\i}s Vasco UPV/EHU, \\
Apartado 644,  48080 Bilbao, Spain}

\emailAdd{emegias@ifae.es}
\emailAdd{manuel.valle@ehu.es}

\abstract{We compute the partition function for non-interacting chiral
  fermions at second order in a derivative expansion of an arbitrary
  time-independent gravitational and gauge background.  We find that
  Pauli-Villars regularization of the vacuum part is needed to get
  consistent results. We use our results to discuss some features of
  the non-dissipative constitutive relations of second order
  hydrodynamics.}

\begin{document}
\maketitle
\flushbottom


\section{Introduction}

One of the most fruitful techniques to study physical systems out of
equilibrium is the hydrodynamical approach, in which it is assumed that
the scales of variation of its observables are much longer than any
microphysical scale in the system (see e.g.~\cite{Kovtun:2012rj} for a
review).  The key ingredients to study the hydrodynamical systems are
the so-called constitutive relations, which are expressions relating
the conserved currents of the systems, energy-momentum tensor and
charged currents, with fluid variables like temperature, chemical
potential, and fluid velocity.  The hydrodynamical approach organizes
the constitutive relations in a derivative expansion of the fluid
variables, and the various terms appearing in this expansion are
multiplied by transport coefficients or susceptibilities.  Some of
these coefficients are responsible for dissipative effects, as they
induce an entropy production in the system out of equilibrium.
Examples of dissipative coefficients at first order in the
hydrodynamical expansion are the shear viscosity $\eta$ and bulk
viscosity~$\zeta$~\cite{Kovtun:2012rj,Arnold:2006fz}.  Other kind of
coefficients in the constitutive relations is related to the static
response of the system to an external perturbation, and they can be
obtained from the equilibrium properties. The magnetic susceptibility
pertains to this kind.

In the past few years a new set of transport coefficients induced by
chiral anomalies has received much attention and interest.  In
presence of anomalies the currents are no longer conserved, and this
has important effects in the constitutive relations.  Some examples of
anomalous coefficients at first order are the {\it chiral magnetic
  conductivity}, which is responsible for the generation of an
electric current parallel to a strong magnetic field in the
system~\cite{Fukushima:2008xe}, and the {\it chiral vortical
  conductivity}, in which the electric current is induced by a
vortex~\cite{Son:2009tf}.  These conductivities are almost
completely fixed by imposing the requirement of zero entropy
production in the equation for the divergence of the entropy
current. 

Recently, it has been shown in~\cite{Banerjee:2012iz,Jensen:2012jh}
that it is not necessary to resort to entropy arguments to obtain the
non-dissipative part of the anomalous constitutive relations.  The
existence of a local partition function that reproduces the consistent
currents in stationary conditions is all that is needed.  It turns out
that the determination of the most general partition function in a
stationary background becomes an important issue not only with regard
to the thermodynamics, but also with hydrodynamics 
(see~\cite{Bhattacharyya:2013lha} for considerations concerning 
the construction of an entropy current from the partition function). 
Other methods to compute the transport coefficients from a microscopic theory include
kinetic theory~\cite{Jeon:1995zm, Arnold:2000dr, Arnold:2002zm}, Kubo
formulae~\cite{Landsteiner:2012kd} and fluid/gravity
correspondence~\cite{Bhattacharyya:2008jc}. In particular, the second
method allowed the identification of a purely temperature dependent contribution
in the chiral vortical conductivity not determined by the second law
of thermodynamics, and it was shown to arise when the system features
a mixed gauge-gravitational
anomaly~\cite{Landsteiner:2011cp,Landsteiner:2011iq}. This was later
confirmed by other
methods~\cite{Chapman:2012my,Valle:2012em,Jensen:2012kj,Megias:2013joa,Banerjee:2013fqa}.

In order to gain more insight into the effects of anomalies, it may be
of interest to go to higher orders in the hydrodynamical derivative
expansion. In this paper we do this by considering the manageable
problem of an ideal fluid of chiral fermions.  The main goal is the
computation of the partition function at second order in the
derivative expansion for this system. A classification of terms
contributing to this order in the constitutive relations was done
in~\cite{Kharzeev:2011ds}.  Following~\cite{Banerjee:2012iz}, we have
considered an arbitrary time-independent background given by the line
element and U(1) gauge connection
\begin{equation}
\begin{split}
ds^2 &= G_{\mu\nu} dx^\mu dx^\nu = -e^{2\sigma(\bm{x})} (dt+a_i(\bm{  x}) dx^i)^2 +g_{ij}(\bm{x}) dx^idx^j \,,  \\ 
\A_\mu &= (A_0(\bm{x}), \bm{\A}(\bm{x})) \,. 
\end{split}
\end{equation}
It is convenient to introduce the combination $A_i \equiv \A_i -
A_0a_i$, which is invariant under the Kaluza-Klein gauge
transformation given by the time reparametrization $t \to t +
\phi(\bm{x})$, $\bm{x} \to \bm{x}$. The most general parity even
partition function to second order in the derivative expansion is
built from seven scalar and two pseudo-scalar quantities as
follows~\cite{Bhattacharyya:2013ida, Bhattacharyya:2014bha}
\begin{equation}
\begin{split}
\cW_2 &= \int d^3x \sqrt{g} \Big[ M_1(\sigma,A_0) T_0^2 \, e^{-2\sigma} \nabla^i \sigma \nabla_i\sigma + 
\frac{M_2(\sigma,A_0)}{T_0^2} \,  \nabla^i A_0 \nabla_i A_0  \\
& \quad - M_3(\sigma,A_0) \, e^{-\sigma}\,  \nabla^i \sigma \nabla_i A_0 + T_0^2 M_4(\sigma,A_0)  f_{ij}f^{ij} + M_5(\sigma,A_0)  F_{ij}F^{ij}  \\
& \quad + T_0 M_6(\sigma,A_0)  f_{ij}F^{ij} + M_7(\sigma,A_0) R \Big]  \\ 
&\quad + \int d^3x \sqrt{g} \Big[ N_1(\sigma, A_0)  \epsilon^{i j k}  \partial_i A_0 f_{j k} + 
      N_2(\sigma, A_0)  \epsilon^{i j k}  \partial_i A_0 F_{j k}    \Big] \, ,  \label{eq:W2}
\end{split}
\end{equation}
where $T_0^{-1}$ is the period of the imaginary time, $R$ is the
three-dimensional Ricci scalar from $g_{i j}$, and we have defined the
strength tensors $F_{ij} = \partial_i A_j - \partial_j A_i$ and
$f_{ij} = \partial_i a_j - \partial_j a_i$.  The functions
$M_i(\sigma,A_0)$, $N_j(\sigma, A_0)$, $i = 1,\dots,7$, $j = 1 , 2$,
depend on the specific system.  Here, we have computed them for an
ideal gas of Weyl and Dirac fermions.  We find that the parity odd
part of the partition function parametrized by  the $N_j$ vanishes in the absence of time reversal symmetry breaking. 
After using Pauli-Villars regularization, we identify the terms in the partition function related to the trace anomaly.  
We also obtain partially the form of non-dissipative constitutive relations in the Landau 
frame by keeping only the terms of second derivatives of the fluid and background fields.

The manuscript is organized as follows. In
section~\ref{sec:current_emt} we obtain the expressions of the charged
$U(1)$ current and energy-momentum tensor for a theory of free Dirac
fermions in $3+1$ dimensions in terms of the thermal Green
function. Then, we compute in section~\ref{sec:green_function} the
Green function up to second order in derivatives of the background
fields. With these results we obtain in section~\ref{sec:1st_order}
the equilibrium expectation values of the current and energy-momentum
tensor as well as the partition function at first order, and in
section~\ref{sec:2nd_order} the energy and charge density at second
order. We present in section~\ref{sec:Partition_func_2ndorder} our
result for the equilibrium partition function at second order, and
this is used in section~\ref{sec:constitutive_relations} to compute
the non-dissipative part of the second order constitutive relations in
the parity even sector. Finally we conclude with a discussion of our
results in section~\ref{sec:Conclusions}.

\section{Theory of free Dirac fermions: $U(1)$ current and energy-momentum tensor}
\label{sec:current_emt}

Our main goal in this manuscript is to determine the partition
function of free chiral fermions in $3+1$ dimensions up to
second order in the derivatives of the metric and gauge background
fields.  However, as we explain in section~\ref{subsec:PauliVillars},
we will derive the result by using Pauli-Villars regularization, and
this demands the consideration of the massive theory for the vacuum
contribution. So, for the sake of completeness we develop in
sections~\ref{sec:current_emt} and \ref{sec:green_function} the
formalism for a massive Dirac field. We keep in
appendix~\ref{sec:Dirac_fermions} some technical details of this
formalism.

Since the partition function will be computed from the equilibrium values of the U(1) current and the stress tensor, 
we begin with the expressions
\begin{equation}
\begin{split}
\label{eq:first}
J^\mu &= -\bar\Psi\underline\gamma^\mu  \Psi  \,,  \\ 
T_{\mu \nu} &=  \frac{i}{4} \bar{\Psi} \left[ 
\underline\gamma_\mu \overrightarrow{\nabla}_\nu  - \overleftarrow{\nabla}_\nu \underline\gamma_\mu + 
( \mu \leftrightarrow \nu) \right ] \Psi \,, 
\end{split}
\end{equation}
where it has been assumed that the spinor field satisfies the Dirac equation. 
The left and right currents are defined by $J^\mu_{L,R} = - \bar\Psi\underline\gamma^\mu \cP_{L,R} \Psi$, 
where $\cP_{L,R} = \frac{1}{2}\left( 1 \pm \gamma_5\right)$ are the chiral projectors. 
Using the explicit form of the background 
we have
\begin{align}
J_0 &= - e^{-\sigma} \psi^\dagger \psi \,, \label{eq:cJ0} \\
J^i &= - \psi^\dagger \sigma_i \psi \,, \label{eq:cJi} \\
T_{00} &= \frac{i}{2} e^{\sigma} \left( \psi^\dagger \partial_t \psi - \partial_t\psi^\dagger \psi \right) 
+ e^\sigma A_0 \psi^\dagger \psi 
- \frac{1}{4} e^{3\sigma} \epsilon^{ijk} \partial_j a_k \psi^\dagger {\bf \sigma}_i \psi \,, \label{eq:cT00} \\
T_0^{\;i} &= \frac{i}{4} e^{\sigma} \left( \psi^\dagger \partial_i \psi - \partial_i\psi^\dagger \psi \right)  - 
        \frac{i}{4} e^{\sigma} a_i \left( \psi^\dagger \partial_t \psi - \partial_t\psi^\dagger \psi \right) 
+ \frac{i}{4} \left( \psi^\dagger \sigma_i \partial_t \psi - \partial_t\psi^\dagger \sigma_i \psi \right) \nonumber \\
& \quad + \frac{1}{2} e^\sigma (\A_i - a_i A_0) \psi^\dagger \psi -\frac{1}{8} e^{2\sigma}  \epsilon^{ijk} \partial_j a_k \psi^\dagger \psi
+ \frac{1}{4} e^{\sigma} \epsilon^{ijk} \partial_j \sigma \psi^\dagger \sigma_k \psi + \frac{1}{2}A_0 \psi^\dagger \sigma_i \psi \,, \label{eq:cT0i}
\end{align}
where eqs.~(\ref{eq:cJ0}) and (\ref{eq:cJi}) correspond to the left-handed current, and $\psi$ is the two-component Weyl fermion $\psi_L$, cf. eq.~(\ref{eq:Psi}). 
Similarly, eqs.~(\ref{eq:cT00}) and (\ref{eq:cT0i}) are the contributions to the stress tensor from the left-handed part. The inclusion of right-handed fermions in these formulas is straightforward. We omit in the following the subindex $L$ to simplify the notation.

The expectation values of these quantities at equilibrium may be computed from the thermal Green's function defined as 
\begin{equation}
\langle T\psi(-i\tau,  \bm  x)\psi^\dagger(0, \bm  x') \rangle_\beta=
T_0\sum_n e^{-i\omega_n\tau} \mathcal{G}( \bm  x,  \bm  x',\omega_n)\ ,
\end{equation}
where $\omega_n = \frac{2\pi}{\beta}\left( n + \frac{1}{2}\right)$ are the fermionic Matsubara frequencies and $\beta = 1/T_0$. The precise form of these is
\begin{align}
\langle J_0 \rangle &= T_0\sum_n\left[-e^\sigma \mathrm{tr}\, \cG(\bm  x, \bm  x, \omega_n)\right] \,,  \label{eq:evJ0} \\
\langle J^i \rangle &= -T_0\sum_n \mathrm{tr}\left[ \sigma_i\,\cG(\bm  x, \bm  x, \omega_n)\right] \,, \label{eq:evJi}
\end{align}
and
\begin{align}
\langle T_{00} \rangle &= T_0 \sum_n \left[ e^\sigma (i \omega_n + A_0)\,  \text{tr}\, \cG(\bm  x, \bm  x, \omega_n) - 
    \frac{1}{4} e^{3 \sigma}  \epsilon^{ijk} \partial_j a_k\,\text{tr}\left[\sigma_i\, \cG(\bm  x,\bm  x, \omega_n)\right]  \right]\,,  \label{eq:evT00} \\ 
\langle T_{0}^{\;i} \rangle &= T_0 \sum_n \left[ \frac{i}{4} e^\sigma   \text{tr}\left( \frac{\partial}{\partial x^i}\cG(\bm  x, \bm  x^\prime, \omega_n)  
   -\frac{\partial}{\partial x^{\prime i}}\cG(\bm  x, \bm x^\prime, \omega_n) \right)  \right] \biggl\lvert_{\bm  x' = \bm  x}  \nonumber \\ 
   &\quad +  T_0 \sum_n\left[  \frac{1}{2} e^\sigma A_i\,  \text{tr}\, \cG(\bm  x, \bm  x, \omega_n) 
    + \frac{1}{4} e^{\sigma} \epsilon^{ijk} \partial_j \sigma \,\text{tr} \left[ \sigma_k \cG(\bm  x,\bm  x, \omega_n) \right] \right] \nonumber  \\ 
   &\quad + T_0 \sum_n \left[\frac{1}{2} (i \omega_n + A_0)\,  \text{tr}\left[ \sigma_i \cG(\bm  x, \bm  x, \omega_n) \right] 
   -\frac{1}{2} e^\sigma a_i (i \omega_n + A_0)\,  \text{tr}\, \cG(\bm  x, \bm  x, \omega_n)\right]  \nonumber \\
   &\quad + T_0 \sum_n \left[ -\frac{1}{8} e^{2 \sigma} \epsilon^{ijk} \partial_j a_k\,  \text{tr}\, \cG(\bm  x, \bm  x, \omega_n)\right] \,.  \label{eq:evT0i}
\end{align}
Therefore, the partition function $\cW$ may be determined by integration of the 
variational formulae~\cite{Banerjee:2012iz}
\begin{align}
\langle J^i \rangle &=\frac{T_0}{\sqrt{-G}}\frac{\delta \mathcal{W}}{\delta A_i} \,, & 
\langle J_0 \rangle &=-\frac{T_0 e^{2\sigma}}{\sqrt{-G}}\frac{\delta \mathcal{W}}{\delta A_0} \,,  \label{eq:variationalJ}\\
\langle T_0^{\;i} \rangle &=\frac{T_0}{\sqrt{-G}}\left(\frac{\delta \mathcal{W}}{\delta a_i}-
\!\!A_0 \frac{\delta \cW}{\delta A_i}\right) \, ,&  \langle T_{00} \rangle &=-\frac{T_0 e^{2\sigma}}{\sqrt{-G}}\frac{\delta \cW}{\delta \sigma} \,. \label{eq:variationalT}
\end{align}
Note that a variation of $\cW_2$ in eq.~(\ref {eq:W2}) with respect to
$\sigma$ or $A_0$ always produces terms which are a product of two
first order derivatives. Thus, in order to obtain the form of the
coefficients $M_i(\sigma, A_0)$ it is sufficient to determine such
bilinear contributions in $\langle T_{00} \rangle$ and $\langle
J_0\rangle$.  Clearly, the first six coefficients may be computed by
setting the metric flat, $g_{ij} = \delta_{i j}$, but the
determination of $M_7$ demands the computation of the energy density
to first order in the three-dimensional curvature $R$.

In the next section we will compute the Green function as an expansion
in derivatives of the background fields. After that, we will use the
expressions above to compute the thermal expectation value of the charged current and energy-momentum tensor at equilibrium at first order, and the charge and energy density at second order.


\section{The Green function}
\label{sec:green_function}

There are several ways to compute the two-point Green function. 
We will follow the procedure of ref.~\cite{Manes:2013kka}. 
We can rewrite the action as
\begin{equation}
S = - \int d^4x \sqrt{-G} \, \bar\Psi \underline\gamma^0 \left[i\partial_t - \cH \right] \Psi \,,
\end{equation}
with the Hamiltonian 
\begin{equation}
\cH = -i\left (\frac{1}{4} \omega_0^{\;\; a  b}\gamma_{ab} - i A_0 \right) -
\frac{i}{g^{00}}\, \underline\gamma^0  \left( \underline\gamma^k\,\nabla_k - m \right) \,.
\end{equation}
After rotating to imaginary time $t \to -i\tau$, the Green function satisfies the differential equation
\begin{equation}
-\sqrt{-G}\,\gamma^0 \underline\gamma^0(i\omega_n- \cH)\cG (\bm  x,  \bm  x^\prime,\omega_n)= \delta(\bm  x- \bm x^\prime)\,.
\end{equation}
The Hamiltonian does not depend on terms beyond first order in derivatives of the background fields. 
After some algebra one gets the exact equation for the Green function
\begin{equation}
\Bigl[ (\mathbb{1}_{4\times4} + e^{\sigma(\bm{x})} \gamma^0 \bm{\gamma} \cdot \bm{a}(\bm{x})) i \omega_n - H(\bm{x}) \Bigr] \cG(\bm{x}, \bm{x^\prime} , \omega_n) =  \delta^{(3)} (\bm{x}-\bm{x^\prime} ) \,, \label{eq:eqGreen}
\end{equation}
where
$H(\bm{x}) = H_0(\bm{x}) + H_1(\bm{x})$ with
\begin{align}
H_0 &=  -A_0\mathbb{1}_{4\times4} - i e^\sigma m \gamma^0  
   + e^\sigma \gamma^0 \bm{\gamma} \cdot ( i  \bm{\partial}  + \bm{\A} - A_0 \bm{a}) \,, \\
H_1 &= \frac{i}{2} e^\sigma \gamma^0 \bm{\gamma} \cdot \bm{\partial} \sigma +  
\frac{i}{4} e^{2 \sigma} \gamma^{j k} \partial_j a_k \, ,
\end{align}
and we have defined $\gamma^{jk} = \frac{1}{2}[\gamma^j,\gamma^k]$.
Note that eq.~(\ref{eq:eqGreen}) is Kaluza-Klein gauge invariant, as
it depends on the combination $\A_i - A_0 a_i$.  In addition, the term
proportional to $ie^\sigma \gamma^0 \bm{\gamma} \cdot \bm{\partial} $
of $H_0$ in combination with the term $i e^\sigma \gamma^0 \bm{\gamma}
\cdot \bm{\partial}$ of $H_1$ guarantees the hermiticity of the
operator~$H$.

This equation can be solved order by order in a derivative expansion
of the background fields.  The solution for the Green function will be
of the form $\cG = \cG_0 + \cG_1 + \cG_2 + \dots$, where the subscript
indicates the order in derivatives.

\subsection{Green function at leading order}
\label{subsec:green_zero}

The Green function at leading order is obtained by neglecting $H_1$ in eq.~(\ref{eq:eqGreen}) and evaluating the background fields at a reference point $\bm{z}$. After Fourier transforming this equation, one can solve it easily as explained in ref.~\cite{Manes:2013kka}. The result is
\begin{equation}
\begin{split}
\cG_0(\bm x, \bm x^\prime, \omega_n) &= -\frac{i}{16\pi^{3/2}} e^{i \left(\bm{\A} - (A_0 + i\omega_n)\bm{a} \right)\cdot (\bm{x}-\bm{x^\prime}) - 2\sigma} \int_0^\infty \frac{ds}{s^{5/2}} e^{-\frac{|\bm{x} - \bm{x}^\prime|}{4s} + b^2 s}  \\
&\quad \times \Bigl( -2s\left[ (A_0 + i\omega_n) \mathbb{1}_{4\times4} - i m e^\sigma \gamma^0 \right] +
 i e^\sigma \gamma^0 \gamma^i (x^i - x^{\prime i})   \Bigr)  \,,
 \end{split}
\end{equation}
where
\begin{equation}
b^2 = -m^2 + e^{-2\sigma}\left( A_0 + i\omega_n\right)^2 \,.
\end{equation}
For computational convenience in what follows, we have made use of the
proper time representation.  This allows to transform the integrals in
the space coordinates $\bm{x}$ and $\bm{x}^\prime$, into Gaussian
integrals that are much more analytically treatable.

\subsection{Green function at higher derivative orders}
\label{subsec:green_higher}

We will study next the solution of eq.~(\ref{eq:eqGreen}) at first and second order in the derivative expansion. 
We consider an expansion of the background fields around the reference point~$\bm{z}$, i.e.
\begin{equation}
\Xi(\bm  x)  = \Xi(\bm  z) +(x^i-z^i)\partial_i\Xi(\bm  z) + 
\frac{1}{2} (x^i - z^i)(x^j-z^j)\partial_i \partial_j\Xi(\bm z) + \ldots \,,
\end{equation}
where $\Xi \equiv \sigma \,, A_0\,, \A_k \,, a_k$. 
Then $H(x)$  has the following expansion
\begin{equation}
H(\bm{x}) = H_0(\bm{z}) + \delta_1 H(\bm{x}) + \delta_2 H(\bm{x}) + \ldots \,,
\end{equation}
where the first and second derivative contributions are, respectively,
\begin{align}
\delta_1 H(\bm  x) &= (x^i-z^i)\left.\partial_i H_0\right|_z+H_1(\bm  z)\, ,  \label{eq:d1H} \\
\delta_2 H(\bm x) &= \frac{1}{2}(x^i-z^i)(x^j-z^j)\left.\partial_i\partial_j H_0\right|_z + (x^i-z^i)\left.\partial_i H_1\right|_z \,.  \label{eq:d2H}
\end{align}
The expansion of the factor $e^{\sigma(\bm{x})} \gamma^0 \bm{\gamma} \cdot \bm{a}(\bm{x})$ of eq.~(\ref{eq:eqGreen}) up to second order is $e^{\sigma(\bm{x})} \gamma^0 \bm{\gamma} \cdot \bm{a}(\bm{x}) =  
e^{\sigma(\bm{z})} \gamma^0 \bm{\gamma} \cdot \bm{a}(\bm{z}) + \delta_1f(\bm{x}) + \delta_2f(\bm{x})$, with
\begin{align}
\delta_1f(\bm{x}) &= (x^i-z^i)e^{\sigma(\bm{z})} \gamma^0 \bm{\gamma} \cdot \big(\partial_i \bm{a}(\bm{z}) + \bm{a}(\bm{z})\partial_i\sigma(\bm{z}) \big) \,,  \label{eq:d1f} \\
\delta_2f(\bm{x}) &= \frac{1}{2}(x^i-z^i)(x^j-z^j) e^{\sigma(\bm{z})} \gamma^0 \bm{\gamma} \cdot \Big( 2 \partial_i\bm{a}(\bm{z})\partial_j\sigma(\bm{z}) + \partial_i\partial_j\bm{a}(\bm{z})  \nonumber \\
&\quad +  \bm{a}(\bm{z})\partial_i\sigma(\bm{z})\partial_j\sigma(\bm{z}) + \bm{a}(\bm{z})\partial_i\partial_j\sigma(\bm{z})  \Big)  \,. \label{eq:d2f}
\end{align}
Note that the background fields in these expansions are always evaluated at the reference point $\bm{z}$.
 Substituting the expansions eqs.~(\ref{eq:d1H})-(\ref{eq:d2f}) into eq.~(\ref{eq:eqGreen}) yields the following 
 differential equations for  $\cG_1$ and $\cG_2$ respectively,
\begin{align}
\big( i\omega_n - H_0(z)\big)\cG_1(\bm x, \bm x^\prime, \omega_n)  &=  \big( \delta_1 H(\bm{x})   - \delta_1 f(\bm{x})i\omega_n    \big) \cG_0(\bm x, \bm x^\prime, \omega_n)  \,,  \\
\big( i\omega_n - H_0(z)\big)\cG_2(\bm x, \bm x^\prime, \omega_n)  &=  \big( \delta_2 H(\bm{x})   - \delta_2 f(\bm{x})i\omega_n    \big) \cG_0(\bm x, \bm x^\prime, \omega_n)  \nonumber \\
& \quad +  \big( \delta_1 H(\bm{x})   - \delta_1 f(\bm{x})i\omega_n    \big) \cG_1(\bm x, \bm x^\prime, \omega_n)  \,.
\end{align}
The solution of the Green function at second order, $\cG_2$,  is more involved than at first order, 
and demands the computation of direct and exchange terms. These equations can be solved in a Schwinger-Dyson expansion to get
\begin{align}
\cG_1( \bm  x,  \bm  x^\prime , \omega_n) &= \int d^3 x^{\prime\prime}\, \cG_0( \bm  x,  \bm  x^{\prime\prime},\omega_n) \bigl(\delta_1 H(\bm  x^{\prime\prime}) - \delta_1 f(\bm x^{\prime\prime} )i\omega_n \bigr)\cG_0( \bm  x^{\prime\prime},  \bm  x^{\prime},\omega_n) \,, \\
\cG_2( \bm  x,  \bm  x^\prime , \omega_n) &= \int d^3 x^{\prime\prime}\, \cG_0( \bm  x,  \bm  x^{\prime\prime},\omega_n) \bigl(\delta_2 H(\bm  x^{\prime\prime})  - \delta_2 f(\bm x^{\prime\prime} )i\omega_n  \bigr)\cG_0( \bm  x^{\prime\prime},  \bm  x^{\prime},\omega_n) \nonumber \\
&\quad +\int d^3 x^{\prime\prime}\, \cG_0( \bm  x,  \bm  x^{\prime\prime},\omega_n) \bigl(\delta_1 H(\bm  x^{\prime\prime}) - \delta_1 f(\bm  x^{\prime\prime} )i\omega_n  \bigr)\cG_1 ( \bm  x^{\prime\prime},  \bm  x^{\prime},\omega_n)  \,.
\end{align}
The evaluation of these integrals is rather lengthy, specially those for the second order Green function. 
Each of these integrals involves the product of two Green's functions,
 and requires the integration over  two proper times $\int_0^\infty ds_1 \int_0^\infty ds_2$. 
 The best way to proceed is to work with new variables $\rho \equiv s_1 + s_2$, $s_1 \equiv \rho \, \xi$, so that the double integral in proper times becomes
\begin{equation}
\int_0^\infty ds_1 \int_0^\infty ds_2 \, f(s_1,s_2) = \int_0^\infty d\rho \, \rho \int_0^1 d\xi \, f(\rho \xi, \rho(1-\xi)) \,.
\end{equation}
The integrals in $\xi$ are finite and can be done straightforwardly in general, so that one ends up with expressions which have to be integrated in the parameter $\rho$. The possible appearance of divergences in the integral over $\rho$ and its regularization 
will be  explained in detail in section~\ref{sec:2nd_order}. 
The complete expressions for~$\cG_1$ and~$\cG_2$ are very lengthy and will not be presented here. Instead, we will use them in the next two sections to compute the thermal expectation values of the current and energy-momentum tensor at first and second order in derivatives.


\section{Covariant current and stress tensor at first order}
\label{sec:1st_order}

After obtaining the thermal Green function at first order, we can compute the $U(1)$ current and energy-momentum tensor at this order by using eqs.~(\ref{eq:evJ0})-(\ref{eq:evT0i}). In the following we will restrict ourselves to a theory with one 
left Weyl  fermion. We are focusing on the parity-odd contributions. In order to compute the result at first order in derivatives, we need to evaluate each term of these equations to the appropriate order in the Green function. In particular, the formulas for $\langle J^i\rangle$ and $\langle T_{0}^{\;i} \rangle$ become
\begin{align}
\langle J^i \rangle &= -T_0\sum_n \mathrm{tr}\,\left[ \sigma_i\,\cG_1(\bm  x, \bm  x, \omega_n)\right] \,, \label{eq:1evJi}  \\
\langle T_{0}^{\;i} \rangle &= T_0 \sum_n \left[ \frac{i}{4} e^\sigma   \mathrm{tr}\left( \frac{\partial}{\partial x^i}\cG_1(\bm  x, \bm  x^\prime, \omega_n)  
   -\frac{\partial}{\partial x^{\prime i}}\cG_1(\bm  x, \bm x^\prime, \omega_n) \right)  \right] \biggl\lvert_{\bm  x' = \bm  x}  \nonumber \\ 
   &\quad + T_0 \sum_n\left[  \frac{1}{2} e^\sigma A_i\,  \text{tr}\, \cG_1(\bm  x, \bm  x, \omega_n) 
    + \frac{1}{4} e^{\sigma} \epsilon^{ijk} \partial_j \sigma \,\text{tr} \left[ \sigma_k \cG_0(\bm  x,\bm  x, \omega_n) \right] \right] \nonumber  \\ 
   &\quad + T_0 \sum_n \left[\frac{1}{2} (i \omega_n + A_0)\,  \text{tr}\left[ \sigma_i \cG_1(\bm  x, \bm  x, \omega_n) \right] 
   -\frac{1}{2} e^\sigma a_i (i \omega_n + A_0)\,  \text{tr}\, \cG_1(\bm  x, \bm  x, \omega_n)\right]  \nonumber \\
   &\quad + T_0 \sum_n \left[ -\frac{1}{8} e^{2 \sigma} \epsilon^{ijk} \partial_j a_k\,  \text{tr}\, \cG_0(\bm  x, \bm  x, \omega_n)\right] \,.  \label{eq:1evT0i}
\end{align}
The traces that will be relevant for this computation are
\begin{align}
\mathrm{tr}\, \cG_0(\bm x, \bm x, \omega_n) &= -\frac{e^{-2\sigma}}{4\pi^{3/2}} \int_0^\infty \frac{d\rho}{\rho^{3/2}} e^{b^2\rho} \, \tilde\omega_n \,, \\
\mathrm{tr} \left[ \sigma_i \, \cG_1(\bm x, \bm x, \omega_n) \right] &= \frac{1}{32\pi^{3/2}}  
   \int_0^\infty \frac{d\rho}{\rho^{3/2}} e^{b^2\rho} \epsilon^{ijk} \bigl[ \partial_j a_k - 
    8\rho e^{-2\sigma}(\partial_j A_k + A_0 \partial_j a_k) \tilde\omega_n \nonumber \\
&\quad + 6\rho e^{-2\sigma} \partial_j a_k \,\tilde\omega_n^2  \bigr]  \,,  \label{eq:trsiG1} \\
\mathrm{tr}\biggl( \frac{\partial}{\partial x^i}\cG_1(\bm  x, \bm  x^\prime, \omega_n)  
   &-\frac{\partial}{\partial x^{\prime i}}\cG_1(\bm  x, \bm x^\prime, \omega_n) \biggr)  \biggl\lvert_{\bm  x' = \bm  x} 
= \frac{i e^{-\sigma}}{8\pi^{3/2}}  \int_0^\infty \frac{d\rho}{\rho^{3/2}} e^{b^2\rho} \nonumber \\
&\quad \times \epsilon^{ijk} \bigl[ 2(\partial_j A_k + A_0 \partial_j a_k) - \partial_j a_k \tilde\omega_n \bigr] \,,
\end{align}
where we have defined $\tilde\omega_n \equiv A_0 + i\omega_n $. 
The remaining traces $\mathrm{tr}\, \cG_1(\bm x, \bm x, \omega_n)$ 
and $\mathrm{tr} \left[\sigma_i \, \cG_0(\bm x, \bm x, \omega_n) \right]$ vanish. 
After performing the summation over Matsubara frequencies as explained in appendix~\ref{sec:technical} 
and integrating in the proper time, this leads to the result
\begin{align}
\langle J_0 \rangle_1 &= 0 \,, \\
\langle J^i \rangle_1 &= e^{-\sigma} \epsilon^{ijk} \left[ C A_0\partial_j A_k 
   + \left(\frac{1}{2}C A_0^2 + C_2 T_0^2 \right) \partial_j a_k \right] \,,  \label{eq:janom} \\
\langle T_{00} \rangle_1  &= 0  \,,  \label{eq:O1T00}\\ 
\langle T_0^{\;i} \rangle_1 &= e^{-\sigma} \epsilon^{ijk} \left[ \left(-\frac{C}{2} A_0^2 
   + C_2 T_0^2 \right)\partial_j A_k + \left( -\frac{C}{6}A_0^3 - C_2 T_0^2 A_0 \right)\partial_j a_k \right] \,, \label{eq:O1T0i}
\end{align}
where $A_k = \A_k - A_0 a_k$, and the constants take the values
\begin{equation}
  C = -\frac{1}{4\pi^2} \,, \qquad C_2 = \frac{1}{24} \,. \label{eq:CC2}
\end{equation}
It has been indicated in~\cite{Banerjee:2012iz} the possible appearance of contributions in eqs.~(\ref{eq:janom}) and (\ref{eq:O1T0i}) of the form $\langle J^i \rangle \sim C_0 e^{-\sigma} \epsilon^{ijk} \partial_j A_k$ and $\langle T_0^{\;i} \rangle \sim C_1 e^{-\sigma} \epsilon^{ijk} \partial_j a_k$. These terms violate ${\cal CPT}$ invariance, and our result leads correctly to a vanishing value for $C_0$ and $C_1$.

The method explained in previous sections makes use of Kaluza-Klein and gauge invariant quantities. 
As a consequence, the $U(1)$ current that we obtain is the covariant current. 
It is related to the consistent current by~\cite{Banerjee:2012iz}
\begin{equation}
J^\mu = J_{\textrm{cons}}^\mu - \frac{C}{6} \epsilon^{\mu\nu\alpha\beta}\A_\nu\cF_{\alpha\beta} \,. \label{eq:JcovJcons}
\end{equation}
Note that the difference between consistent and covariant currents,
which is the Bardeen polynomial, is only first order in
derivatives~\cite{Bardeen:1984pm}. Using eq.~(\ref{eq:JcovJcons}), the
result for the consistent current at first order reads
\begin{align}
\langle J_{\textrm{cons},0} \rangle_1 &= -e^\sigma \epsilon^{ijk} \left[  \frac{C}{3} A_i \partial_j A_k + \frac{C}{3} A_0 A_i\partial_j a_k \right] \,,  \label{eq:O1J0}\\
\langle J_{\textrm{cons}}^i \rangle_1 &= e^{-\sigma} \epsilon^{ijk} \left[ \frac{2}{3}C A_0\partial_j A_k + \left(\frac{1}{6}C A_0^2 + C_2 T_0^2 \right) \partial_j a_k + \frac{C}{3}A_k\partial_j A_0  \right] \,. \label{eq:O1Ji}
\end{align}
Now, the general form of the consistent partition function at first order is~\cite{Banerjee:2012iz}
\begin{equation}
\cW_1 = \int d^3x \sqrt{g}  \left[ \alpha_1(\sigma,A_0) \epsilon^{ijk}A_i F_{jk} + \alpha_2(\sigma,A_0) \epsilon^{ijk} A_i f_{jk} + \alpha_3(\sigma,A_0) \epsilon^{ijk} a_i f_{jk}  \right] \,. \label{eq:W1}
\end{equation}
Using this formula in eqs.~(\ref{eq:variationalJ}) and (\ref{eq:variationalT}) and comparing with the results given by eqs.~(\ref{eq:O1T00})-(\ref{eq:O1T0i}) and (\ref{eq:O1J0})-(\ref{eq:O1Ji}), 
one gets the following explicit expressions for the functions~$\alpha_i(\sigma,A_0)$,
\begin{equation}
\alpha_1(\sigma,A_0) =  \frac{C}{6T_0}A_0 \,, \qquad \alpha_2(\sigma,A_0) =  \frac{1}{2}\left( \frac{C}{6T_0}A_0^2 + C_2 T_0 \right) \,, \qquad \alpha_3(\sigma,A_0) = 0 \,.
\end{equation}
The coefficient $\alpha_3(\sigma,A_0)$ is proportional to $C_1$, which is zero in a ${\cal CPT}$ invariant theory as mentioned above.


\section{Energy and charge density at second order. Renormalization}
\label{sec:2nd_order}
 
In this section we will give the parts of $\langle J_0\rangle_2$ and
$\langle T_{00}\rangle_2$ that are bilinear in derivatives of the
background fields, i.e. contributions which are the product of first order
terms, as well as the part proportional to the three-dimensional
curvature.  These expectation values follow from the equilibrium
partition function which, in principle, can include all scalars
containing two space derivatives.  The explicit computation we will
present shows that the four bilinear pseudo-scalars
 \begin{equation}
\epsilon^{i j k} \nabla_i \sigma  f_{j k}\,, \qquad \epsilon^{i j k} \nabla_i \sigma  F_{j k}\,, \qquad \epsilon^{i j k} \nabla_i A_0  f_{j k}\,, \qquad \epsilon^{i j k} \nabla_i A_0  F_{j k}\,, \label{eq:pseudo_scalars}
\end{equation}
are absent. This is remarkable and requires an explanation.  
Under time reversal, the signature of $A_0$ and $\sigma$ is $+1$,
while that of $A_i$ and $a_i$ is $-1$.  As we have seen in the
previous section, the consistent partition function exhibits a parity-odd dependence through
the terms $\epsilon^{i j k} A_i F_{j k}$ and $\epsilon^{i j k} A_i
f_{j k}$.  According to this, the consistent partition function at
first order does not change its sign under time reversal.  On the other hand,
the four pseudo-scalars in~(\ref{eq:pseudo_scalars})
multiplied by any function of $A_0$ and $\sigma$ change  their sign
under ${\cal T}$, and it turns out that the parity violating partition
function at second derivative order behaves in opposite way to that of
first order.  If follows that if the underlying Hamiltonian is
invariant under ${\cal T}$, the parity violating part of the partition
function at second order vanishes.

In the parity even sector the possible terms that can appear at second
order have been classified in~\cite{Bhattacharyya:2014bha}, and they
are written in eq.~(\ref{eq:W2}).  Our goal is to compute explicitly
the coefficients $M_i$, $i=1,\dots,7$, for a free Weyl fermion and a
massless Dirac fermion, in order to ascertain possible differences
regarding the chiral anomaly at second order.

\subsection{$\langle T_{00}\rangle$ and $\langle J_0 \rangle$ for Weyl fermions}

The evaluation of eqs.~(\ref{eq:evJ0}) and (\ref{eq:evT00}) for a
chiral fermion with $\mathcal{G}_2(\bm{x}, \bm{x},\omega_n)$ produces (see appendix~\ref{sec:technical} for details)
\begin{align}
\langle J_0\rangle_2 &=  \frac{1}{24 \pi^2} \biggl(
  -\nabla^i A_0 \nabla_i \sigma + \frac{1}{2}e^{2 \sigma}  f_{i j} F^{ij}  + 
  \frac{1}{2} A_0 e^{2 \sigma}f_{i j} f^{ij} \biggr) \mathcal{N}_\Lambda(\sigma, A_0)  \notag \\ 
& \quad  + \frac{1}{48 \pi^2} \left(
  \nabla^i A_0 \nabla_i A_0 + \frac{e^{2 \sigma}}{2} A_0^2 f_{i j} f^{ij}  + \frac{e^{2 \sigma}}{2} F_{i j} F^{ij} 
 +e^{2 \sigma} A_0 f_{i j} F^{ij} \right) \frac{\partial \mathcal{N}_\Lambda}{ \partial A_0}  \notag \\ 
& \quad  -\frac{1}{24\pi^2} A_0 \nabla^i \sigma \nabla_i \sigma + \frac{7}{96\pi^2} \nabla^i A_0 \nabla_i \sigma +
 \frac{5}{192\pi^2} e^{2\sigma} f_{ij} F^{ij} \notag \\
 & \quad  +  \frac{3}{64\pi^2} e^{2\sigma} A_0 f_{ij}f^{ij} +\frac{A_0}{48\pi^2} R   \,, \label{eq:J0} 
\end{align}
\begin{align}
\langle T_{00}\rangle_2 &=  \frac{1}{48 \pi^2} \biggl(
  \nabla^i A_0 \nabla_i A_0 + \frac{e^{2 \sigma}}{2} A_0^2 f_{i j} f^{ij}  + \frac{e^{2 \sigma}}{2} F_{i j} F^{ij} 
 +e^{2 \sigma} A_0 f_{i j} F^{ij} \biggr) \mathcal{N}_\Lambda(\sigma, A_0) \notag \\ 
& \quad  +\biggl(\frac{A_0^2}{48\pi^2} + \frac{T_0^2}{144} \biggr) \nabla^i \sigma \nabla_i \sigma 
- \frac{A_0}{12\pi^2} \nabla^i A_0 \nabla_i \sigma + \frac{5 e^{2 \sigma}}{64 \pi^2}  A_0 f_{ij} F^{i j} \notag \\
& \quad + \frac{7}{384 \pi^2} \biggl(2  \nabla^i A_0 \nabla_i A_0 + e^{2 \sigma} F_{i j} F^{i j} \biggr) +
 \biggl(\frac{19 A_0^2}{384 \pi^2} -\frac{T_0^2}{288} \biggr) e^{2 \sigma} f_{ij} f^{ij}  \notag \\
 & \quad -\biggl(\frac{A_0^2}{96\pi^2} + \frac{T_0^2}{288} \biggr) R -
 \frac{23 e^{4 \sigma}}{3072 \pi^2}  f_{i j} f^{i j}  \int_{1/\Lambda^2}^\infty \frac{d \rho}{\rho^2} + 
 \frac{1}{4} e^{3 \sigma} \text{rot}\,  \bm{a} \cdot  
  \langle \tilde{\bm{J}}\rangle_1  \,, \label{eq:T00}  
\end{align}
where $\mathcal{N}_\Lambda(\sigma, A_0)$ turns out to be the following 
combination that includes  vacuum and thermal effects in the massless case 
\begin{equation}
\mathcal{N}_\Lambda(\sigma, A_0) \equiv   \int_{1/\Lambda^2}^{1/\Lambda_{IR}^2}  \frac{d\rho}{\rho} + 
2 \sum_{n=1}^\infty \int_0^{1/\Lambda_{IR}^2} \exp \biggl(-\frac{e^{2 \sigma} n^2}{4 T_0^2 \rho} \biggr) 
\cos \bigl(n (\pi  - A_0/T_0) \bigr) \, . 
\end{equation} 
Note that the replacement $\Lambda \to \infty$ in the second integral is
safe.  We should note that both integrals  are separately
infrared divergent, but the summation in the thermal part removes the
dependence on the IR regulator $\Lambda_{IR} \to 0$, so the leading
logarithmic dependence of $\mathcal{N}_\Lambda$ is $\ln (e^{2 \sigma}
\Lambda^2/T_0^2)$.  A simple computation leads to
\begin{equation}
\mathcal{N}_\Lambda(\sigma, A_0) =  \ln \frac{e^{2 \sigma} \Lambda^2}{T_0^2}+ \gamma_E - 2 \ln2  + Q\left(\frac{A_0}{T_0}\right), 
\end{equation}
where $Q(\nu)$ is the analytic continuation of the series 
\begin{equation}
Q(\nu) = -2\sum_{n=1}^\infty (-1)^n\cosh(n\nu)\log(n^2) \,.
\end{equation}
Hence, $\partial \mathcal{N}_\Lambda/\partial A_0 = T_0^{-1} Q'(\nu)$,
where $\nu = A_0/T_0$. Note that, although the last term of
eq.~(\ref{eq:T00}) comes from the anomalous current of
eq.~(\ref{eq:janom}), it produces an even parity contribution
proportional to a combination of $f_{i j} f^{i j}$ and $f_{i j} F^{i
  j}$.  Since other terms with this parametric dependence are already
present in $T_{00}$, the contribution of the chiral anomaly at second
order appears mixed with other parity even terms such as the
coefficient of $\mathcal{N}_\Lambda$, which corresponds to the trace
anomaly as we will see later.

\subsection{Vacuum expectation values  from regulators: Pauli-Villars regularization}
\label{subsec:PauliVillars}

The vacuum contribution to the thermal expectation values is
logarithmically divergent in the UV, so we need to choose a
regularization procedure. This can be done in a gauge invariant way by
means of Pauli-Villars regularization.  In our case, it suffices to
consider three heavy fermions with masses $M_\ell$ and weights
$C_\ell$ obeying the conditions
\begin{equation}
\begin{split}
1 + \sum_{\ell=1}^3 C_\ell &= 0 \, , \\ 
\sum_{\ell=1}^3 C_\ell M_\ell^2 & = 0 \, .  \label{eq:PVcond}
\end{split}
\end{equation}
A simple choice satisfying these constraints is $C_1=1$, $C_2 = C_3 = -1$, and $M_1 = \sqrt{2} M$, $M_2 = M_3 = M$, where $M$ is a large mass. 

For a  massive fermion $\Psi_\ell$, the vacuum expectation values that result from eq.~(\ref{eq:first}) by  
projecting on the left component read
\begin{align}
\langle J_0\rangle_2^\mathrm{vac} &=  \frac{1}{24 \pi^2} \biggl(
  -\nabla^i A_0 \nabla_i \sigma + \frac{1}{2}e^{2 \sigma}  f_{i j} F^{ij}  + 
  \frac{1}{2} A_0 e^{2 \sigma}f_{i j} f^{ij} \biggr) \int_{1/\Lambda^2}^\infty e^{-M_\ell^2 \rho} \frac{d\rho}{\rho}
    \notag \\  
& \quad - \frac{5}{96\pi^2} \nabla^i A_0 \nabla_i \sigma +
\frac{1}{64 \pi^2} \biggl( e^{2 \sigma} f_{i j} F^{i j}  + e^{2 \sigma} A_0 f_{i j} f^{i j}\biggr)  \,, \label{eq:J0PV} \\ 
\langle T_{00} \rangle_2^\mathrm{vac}&=  \biggr[\frac{1}{48 \pi^2} \biggl(
  \nabla^i A_0 \nabla_i A_0 + \frac{e^{2 \sigma}}{2} A_0^2 f_{i j} f^{ij}  + \frac{e^{2 \sigma}}{2} F_{i j} F^{ij}  
+e^{2 \sigma} A_0 f_{i j} F^{ij}  \biggr)  \notag  \\ 
& \quad -\frac{11 M_\ell^2}{3072 \pi^2} e^{4 \sigma} f_{ij} f^{i j} \biggr]  
   \int_{1/\Lambda^2}^\infty e^{-M_\ell^2 \rho} \frac{d\rho}{\rho}  -
   \frac{23}{3072 \pi^2} e^{4 \sigma} f_{i j} f^{i j}  \int_{1/\Lambda^2}^\infty  e^{-M_\ell^2 \rho} \frac{d \rho}{\rho^2}  \notag \\
 & \quad + \frac{M_\ell^2}{64 \pi^2} e^{2 \sigma}  \nabla^i \sigma \nabla_i \sigma +
  \frac{1}{64 \pi^2}  \nabla^i A_0 \nabla_i A_0
  \notag \\
  &\quad 
  - \frac{1}{384 \pi^2}  \biggr( e^{2 \sigma} F_{i j} F^{i j} + 2 e^{2 \sigma} A_0  f_{i j} F^{i j}
   + e^{2 \sigma} A_0^2  f_{i j} f^{i j}\biggl) \notag \\ 
   & \quad + 
   \frac{1}{192 \pi^2} e^{2 \sigma} R \int_{1/\Lambda^2}^\infty  M_\ell^2 e^{-M_\ell^2 \rho} \frac{d \rho}{\rho} \, . \label{eq:T00PV}
 \end{align}  
 In the expression for $\langle T_{00} \rangle_2^\mathrm{vac} $ we
 have not included the vacuum contribution from the term proportional
 to $\langle \Psi^\dagger \gamma^{i j} \Psi \rangle_1^\mathrm{vac}
 f_{i j}$, which after projection on the left component, could combine
 with the last term of eq.~(\ref{eq:T00}) to give a possible finite
 part. However, such a finite part vanishes when the contributions
 from the physical field and the three regulators are combined:
 \begin{equation}
\int_{1/\Lambda^2}^\infty \frac{d \rho} {\rho^2}  + 
\sum_{\ell = 1}^3 C_\ell \int_{1/\Lambda^2}^\infty e^{-M_\ell^2 \rho} \frac{(1 + M_\ell^2 \rho)}{\rho^2} d\rho = 
-\frac{e^{-2 M^2 \rho}(-1 + e^{M^2 \rho})^2}{\rho} \Biggl\lvert_{1/\Lambda^2 \to 0}^\infty = 0 \,. 
 \end{equation}
 The finite terms independent on the mass are generated by the integrals 
 \begin{equation}
 \int_0^\infty e^{-M_\ell^2 \rho} M_\ell^2 d\rho = 1 \,, \qquad  \int_0^\infty e^{-M_\ell^2 \rho} M_\ell^4 \rho \, d\rho = 1 \,.
 \end{equation}
The use of the values of $C_\ell$ and $M_\ell$ given above yields the following combinations of integrals 
 \begin{align}
 \mathcal{N}_\Lambda + \sum_{\ell=1}^3 C_\ell \int_{1/\Lambda^2}^\infty e^{-M_\ell^2 \rho} \frac{d\rho}{\rho} &=
   2 \gamma_E  - 3 \ln 2 + \ln \frac{e^{2 \sigma} M^2}{T_0^2}  +  Q\left(\frac{A_0}{T_0}\right) \,, \\ 
   \sum_{\ell=1}^3 C_\ell \int_{1/\Lambda^2}^\infty  M_\ell^2 e^{-M_\ell^2 \rho} \frac{d\rho}{\rho} &=-2 M^2  \ln2 \,,  \\
 \int_{1/\Lambda^2}^\infty  \frac{d\rho}{\rho^2}  + 
  \sum_{\ell=1}^3 C_\ell \int_{1/\Lambda^2}^\infty e^{-M_\ell^2 \rho} \frac{d\rho}{\rho^2} &=2 M^2  \ln2 \, , 
 \end{align}
which, together with eq.~(\ref{eq:PVcond}), finally produce the total vacuum contribution of the Pauli-Villars regulators
\begin{align}
\langle J_0\rangle_2^\mathrm{PV} &=  \frac{1}{24 \pi^2} \biggl(
  -\nabla^i A_0 \nabla_i \sigma + \frac{1}{2}e^{2 \sigma}  f_{i j} F^{ij}  + 
  \frac{1}{2} A_0 e^{2 \sigma}f_{i j} f^{ij} \biggr) \biggl(\ln \frac{e^{2 \sigma} \bar{M}^2}{T_0^2} + Q - 
  \mathcal{N}_\Lambda \biggr) \notag \\  
 & \quad + \frac{5}{96\pi^2} \nabla^i A_0 \nabla_i \sigma -
\frac{1}{64 \pi^2} \biggl( e^{2 \sigma} f_{i j} F^{i j}  + e^{2 \sigma} A_0 f_{i j} f^{i j}\biggr) \,, \label{eq:J02PV} \\ 
 \langle T_{00} \rangle_2^\mathrm{PV}&=  \frac{1}{48 \pi^2} \biggl(
  \nabla^i A_0 \nabla_i A_0 + \frac{e^{2 \sigma}}{2} A_0^2 f_{i j} f^{ij}  + \frac{e^{2 \sigma}}{2} F_{i j} F^{ij}  
+e^{2 \sigma} A_0 f_{i j} F^{ij}  \biggr)  \notag  \\  
& \quad  \times  \biggl(\ln \frac{e^{2 \sigma} \bar{M}^2}{T_0^2}  + Q - \mathcal{N}_\Lambda \biggr)  +
\frac{23}{3072 \pi^2} e^{4 \sigma} f_{ij} f^{i j}  \int_{1/\Lambda^2}^\infty  \frac{d\rho}{\rho^2} \notag \\ 
&\quad -\frac{M^2 \ln2}{96 \pi^2} e^{2 \sigma} R - \frac{M^2 \ln2}{128 \pi^2} e^{4 \sigma} f_{ij} f^{i j} 
- \frac{1}{64 \pi^2}  \nabla^i A_0 \nabla_i A_0 \notag \\ 
 &\quad +\frac{1}{384 \pi^2}  \biggr( e^{2 \sigma} F_{i j} F^{i j} + 2 e^{2 \sigma} A_0  f_{i j} F^{i j}
   + e^{2 \sigma} A_0^2  f_{i j} f^{i j}\biggl) \,,   \label{eq:T002PV}
\end{align} 
where we have defined the rescaled Pauli-Villars mass $\bar{M} =
2^{-3/2} e^{\gamma_E} M$ to simplify the expressions. 
Thus, a renormalized expectation value is given by
$\langle \mathcal{O} \rangle_2 + \langle \mathcal{O}
\rangle_2^\mathrm{PV}$, where the first summand is given either by
eq.~(\ref{eq:J0}) or~(\ref{eq:T00}).


\section{Partition function at second order }
\label{sec:Partition_func_2ndorder}

The general expression for the partition function at second order can be written as~\cite{Bhattacharyya:2014bha}
\begin{equation}
\begin{split} \label{eq:W}   
\cW_2 &= \int d^3 x \sqrt{g} \Bigl[ M_1 g^{i j} \partial_i T \partial_j T + 
  M_2 g^{i j} \partial_i \nu \partial_j \nu  + 
  M_3 g^{i j} \partial_i \nu \partial_j T   \\ 
  &\quad +T_0^2 M_4 f_{i j} f^{i j} +    M_5 F_{i j} F^{i j} + T_0 M_6 f_{i j} F^{i j} + M_7 R \Bigr] \,,    
\end{split}
\end{equation}
where $M_i = M_i(T,\nu)$, and 
\begin{equation}
 T = T_0 \,e^{-\sigma}\,, \qquad \nu = \frac{A_0}{T_0} \,. 
\end{equation}
Using the variational formulae eqs.~(\ref{eq:variationalJ})-(\ref{eq:variationalT}) with eq.~(\ref{eq:W}), this gives
\begin{align}
 \langle J_0 \rangle_2 &=  T_0 e^{-\sigma} \left( e^\sigma M_3 - 
                                 T_0 \frac{\partial M_1}{\partial \nu}  + 
                                 T_0 \frac{\partial M_3}{\partial T}\right)\nabla^i \sigma \nabla_i\sigma  \notag   \\ 
         &\quad - 2 \frac{\partial M_2}{\partial T}  \nabla^i \sigma \nabla_i A_0  +
                             \frac{e^\sigma}{T_0^2} \frac{\partial M_2}{\partial \nu}  \nabla^i A_0 \nabla_i A_0 - 
                              \frac{\partial M_5}{\partial \nu}  F_{i j} F^{i j}   \notag \\ 
        &\quad - T_0 \frac{\partial M_6}{\partial \nu}  f_{i j} F^{i j}  -  T_0^2 \frac{\partial M_4}{\partial \nu}  f_{i j} f^{i j} -
           e^\sigma \frac{\partial M_7}{\partial \nu}  R \, ,  \label{eq:J0M} \\
\langle T_{00} \rangle_2 &= -T_0^3 e^{-2 \sigma} \left(2 e^\sigma M_1 + 
                                 T_0 \frac{\partial M_1}{\partial T} \right)\nabla^i \sigma \nabla_i\sigma + 
                                 2 T_0^2 e^{-\sigma}  \frac{\partial M_1}{\partial \nu}  \nabla^i \sigma \nabla_i A_0  \notag   \\
       & \quad + \left( \frac{\partial M_2}{\partial T}  - 
                                  \frac{\partial M_3}{\partial \nu} \right) \nabla^i A_0 \nabla_i A_0  + 
                                  T_0^2  \frac{\partial M_5}{\partial T}  F_{i j} F^{i j} +  
                                  T_0^3  \frac{\partial M_6}{\partial T}  f_{i j} F^{i j}   \notag \\
        & \quad + T_0^4  \frac{\partial M_4}{\partial T}  f_{i j} f^{i j}  + T_0^2 \frac{\partial M_7}{\partial T} R \, . \label{eq:T00M} 
\end{align}
By using the renormalized expressions of $\langle J_0 \rangle_2$ and
$\langle T_{00}\rangle_2$ computed in section~\ref{sec:2nd_order} and
plugging them into the lhs of eqs.~(\ref{eq:J0M}) and (\ref{eq:T00M}),
one gets a system of 14 equations and 7 functions of two
arguments. After solving these equations one gets the following result
\begin{align}
M_1(T, \nu) &= -\frac{1}{144} \frac{1}{T} - \frac{1}{48\pi^2} \frac{\nu^2}{T}  \,, \label{eq:M1} \\ 
M_2(T, \nu) &=  \frac{1}{48\pi^2} T \left(\ln \frac{\bar{M}^2}{T^2} + Q(\nu) - \frac{1}{4} -\frac{3}{4} \right) \,,  \label{eq:M2} \\ 
M_3(T, \nu) &= - \frac{1}{12\pi^2} \nu \,,  \label{eq:M3}  \\ 
M_4(T, \nu) &= -\frac{1}{96\pi^2} \frac{\nu^2}{T} \left(\ln \frac{\bar{M}^2}{T^2} + Q(\nu) + \frac{11}{4} + 6 \pi^2 C  + \frac{1}{4}\right) + 
 \frac{1}{288}\frac{1}{T}  - \frac{C_2 }{8 T} \ 
 \notag  \\
& \quad + \frac{1}{384\pi^2} \frac{1}{T^3}  M^2 \ln 2 \,, \label{eq:M4}  \\ 
M_5(T, \nu) &=  -\frac{1}{96\pi^2} \frac{1}{T}  \left(\ln \frac{\bar{M}^2}{T^2} + Q(\nu) - \frac{1}{4} + \frac{1}{4} \right) \,, \label{eq:M5} \\
M_6(T, \nu) &=  -\frac{1}{48\pi^2} \frac{\nu}{T} \left(\ln \frac{\bar{M}^2}{T^2} + Q(\nu) + \frac{7}{4} + 6 \pi^2 C  + \frac{1}{4} \right) \,,  \label{eq:M6} \\  
M_7(T, \nu) &= -\frac{1}{288} T - \frac{1}{96\pi^2} T \,  \nu^2  + 
\frac{1}{96 \pi^2} \frac{1}{T} M^2 \ln 2 \,. \label{eq:M7}
\end{align}
The constants $C$ and $C_2$ are given by eq.~(\ref{eq:CC2}), and
$\bar{M}$ is defined after eq.~(\ref{eq:T002PV}). As we will see later the logarithmic dependence in $\bar{M}$ is related to conformal anomalies.  
The combination of terms proportional to $M^2$ in $M_4$ and $M_7$ is a
pure renormalization effect.  These terms can be renormalized by
adding a counterterm proportional to the Ricci scalar $\tilde{R}$ of
the $3+1$ dimensional metric,               
\begin{equation}
\cW_2^{\textrm{ct}} =  -\frac{M^2 \ln 2}{96 \pi^2}  \int d^4x \sqrt{-G} \,  \tilde{R} \,,
\end{equation}
so that the renormalized partition function is $\cW_2^{\textrm{ren}} = \cW_2 + \cW_2^{\textrm{ct}}$. 
By using the relation between the scalar curvatures  
\begin{equation}
\tilde{R} = R + \frac{1}{4} e^{2 \sigma} f_{i j} f^{i j} - 
    \frac{2 e^{-\sigma}}{\sqrt{g}} \partial_i \bigl(g^{ij} \sqrt{g}  e^{\sigma}\partial_j\sigma  \bigr) \,, 
\end{equation}
one gets
\begin{equation}
\begin{split}
\cW_2^{\textrm{ct}} &= -\frac{M^2 \ln 2}{96 \pi^2} \left[ \int d^3x \sqrt{g} \frac{e^\sigma}{T_0} \left( R +
 \frac{1}{4} e^{2\sigma} f_{ij}f^{ij} \right) \right. \\ 
 &\quad  - 2 \left. \int \frac{d^3x}{T_0} \sqrt{g}  \frac{1}{\sqrt{g}} \partial_i\left( g^{ij} \sqrt{g} 
 e^\sigma \partial_j\sigma \right)  \right] \,,  \label{eq:W2ct}
 \end{split}
\end{equation}
where the last term vanishes.  One can see that
this counterterm exactly cancels the $M^2$ terms in $\cW_2$. 
Then the renormalized coefficients $M_{4,7}^{\textrm{ren}}$ are the same as $M_{4,7}$, but removing the terms proportional to $M^2$.

The additive constants in $M_2$, $M_4$, $M_5$ and $M_6$, i.e. $-3/4\,, 1/4\,, 1/4$ and $1/4$, are the finite contributions coming from the Pauli-Villars regulator. We would like to emphasize that these contributions allow the thermal expectation values to be consistent with the partition function result. If they were not taken into account, them the system of equations for the functions $M_i(T,\nu)$ wouldn't have a solution for all the terms considered in $\langle J_0 \rangle$ and $\langle T_{00}\rangle$. 
 It is then remarkable that the vacuum contribution can affect the finite temperature part 
 to make it consistent with the prediction from the partition function.
Of course, for a given coefficient, e.g. $M_5$,
we can remove some of these constants by a redefinition of  $\bar{M}$, 
but  the parametric dependence of the other coefficients on $M_5$ remains unaffected:
\begin{align}
M_2 &= -2 T^2 M_5 - \frac{T}{48\pi^2} \,, \\
M_4^{\textrm{ren}} &= \nu^2 M_5 + \frac{(1-36 C_2)}{288 T} -\frac{(1+2\pi^2 C)\nu^2}{32\pi^2 T} \,, \\
M_6 &= 2\nu M_5 - \frac{(1+3\pi^2 C)\nu}{24\pi^2 T} \,.
\end{align}
Note that there are three independent combinations of $M_{2,4,5,6}$
which do not include logarithmic dependence in $\bar{M}$. They, or a
linear combination of them, will appear in some of the transport
coefficients of the hydrodynamic constitutive relations, see
eq.~(\ref{eq:lambda}). 

To conclude this section, let us examine the
transformation of the partition function given by eq.~(\ref{eq:W}) under a Weyl rescaling.  
The
different quantities involved transform under this as
\begin{equation}
\begin{split}
g_{i j}&\to e^{2 \omega} g_{i j} \,, \qquad   g^{i j} \to e^{-2 \omega} g^{i j}\,,
 \qquad \sigma \to \sigma + \omega \,, \qquad \sqrt{g} \to e^{3 \omega}\sqrt{g} \,,  \\ 
 R & \to e^{-2 \omega} \left(R - 2 g^{i j} \partial_i \omega \partial_j \omega - 4 \nabla^2 \omega \right) \,,
 \end{split}
\end{equation}
and the lower components $A_i$, $\mathcal{A}_i$, as well as  $A_0$, are unchanged. 
Substituting into eq.~(\ref{eq:W}),  and using the formulae   
\begin{equation}
\langle T_{00} \rangle=  -\frac{T_0 e^\sigma}{\sqrt{g}}  \frac{\delta \mathcal{W}_2}{\delta \sigma} \,, \qquad
\langle T^{i j} \rangle =  -\frac{2 T}{\sqrt{g}} g^{i m} g^{j n} \frac{\delta \mathcal{W}_2}{\delta g^{m n}} \,,
\end{equation}
we find the general form of the trace of the stress tensor at equilibrium,  
\begin{equation}
\begin{split}
\label{eq:gentraza}
\frac{1}{\sqrt{g}}\frac{\delta \mathcal{W}_2}{\delta \omega}\biggl|_{\omega =0} &= \frac{1}{T} \Bigl(
    g_{i j} \langle T^{i j} \rangle  - e^{-2 \sigma} \langle T_{00}  \rangle \Bigr)\\
 &= \left(M_1 + T \frac{\partial M_1}{\partial T}   - 4  \frac{\partial^2 M_7^{\textrm{ren}}}{\partial T^2}\right) (\nabla T)^2 \\ 
 & \quad +\left(M_3 + 2 T \frac{\partial M_1}{\partial \nu}   - 8  \frac{\partial^2 M_7^{\textrm{ren}}}{\partial T \, \partial \nu}\right) 
        \nabla^j T\nabla_j \nu \\
  & \quad + \left(T \frac{\partial M_3}{\partial \nu}   - 4  \frac{\partial^2 M_7^{\textrm{ren}}}{\partial \nu^2}\right) (\nabla \nu)^2 + 
   \left(M_7^{\textrm{ren}} - T \frac{\partial M_7^{\textrm{ren}}}{\partial T}\right) R  \\
   & \quad +   \left(T M_3 -4 \frac{\partial M_7^{\textrm{ren}}}{\partial \nu}\right) \nabla^2 \nu + 
   \left(2 T M_1 -4 \frac{\partial M_7^{\textrm{ren}}}{\partial T}\right) \nabla^2 T \\
  &  \quad  +  \left(M_2 - T \frac{\partial M_2}{\partial T}\right) (\nabla \nu)^2 - 
  T_0^2 \left(M_4^{\textrm{ren}} + T \frac{\partial M_4^{\textrm{ren}}}{\partial T}\right) f_{i j} f^{i j}  \\ 
  & \quad   -   \left(M_5 + T \frac{\partial M_5}{\partial T}\right) F_{i j} F^{i j} - 
  T_0 \left(M_6 +  T \frac{\partial M_6}{\partial T}\right) f_{i j} F^{i j}  \, . 
\end{split}
\end{equation}  
The partition function is conformally invariant only if all the coefficients  vanish.
The first four lines in the last equality of eq.~(\ref{eq:gentraza}) only involve $M_1$, $M_3$ and $M_7^{\textrm{ren}}$, 
so  the cancellation of the corresponding coefficients  determines $M_1$ and $M_3$ in terms of  $M_7^{\textrm{ren}}$, 
\begin{equation}
M_1 = \frac{2}{T} \frac{\partial M_7^{\textrm{ren}}}{\partial T} \,  , 
\qquad M_3 = \frac{4}{T} \frac{\partial M_7^{\textrm{ren}}}{\partial \nu} \,  , 
\end{equation}
with $M_7^{\textrm{ren}}(T, \nu) = T  f_7(\nu)$, where $f_i$ is an arbitrary function.   
The remainder conditions  for conformal invariance leads to 
\begin{equation}
M_2 = T f_2(\nu), \qquad M_4^{\textrm{ren}} = T^{-1} f_4(\nu), \qquad M_5 = T^{-1} f_5(\nu),
\qquad M_6 = T^{-1} f_6(\nu) .
\end{equation}

The model considered here only violates conformal invariance because renormalization effects, 
which lead to a logarithmic dependence on $\ln \tfrac{\bar{M}}{T}$ of $M_2$, $M_4^{\textrm{ren}}$, $M_5$ and $M_6$. 
In the case of a free Weyl fermion the anomalous partition function reads 
 \begin{equation}
\begin{split}
\mathcal{W}_\mathrm{anom} &= \frac{1}{24 \pi^2} \int d^3 x \sqrt{g} \frac{1}{T}  \ln \frac{\bar{M}}{T} \\ 
 & \quad \times \left( e^{-2 \sigma}  g^{i j} \partial_i A_0 \partial_j A_0 - \frac{1}{2} A_0^2 f_{i j} f^{ij}  
  -\frac{1}{2} F_{i j} F^{ij} - A_0 f_{i j} F^{ij}\right)  \, ,  
 \end{split}
\end{equation}
which, by using the relation $A_i = \mathcal{A}_i - A_0 a_i$, can be
written only in terms of the four dimensional metric and the field
strength of the gauge field $\mathcal{A}$ as
\begin{equation}
\begin{split}
\mathcal{W}_\mathrm{anom} &= \frac{1}{24 \pi^2} \int d^3 x \sqrt{g} \frac{1}{T}  \ln \frac{\bar{M}}{T} \times \left(-\frac{1}{2} G^{\mu \rho} G^{\nu \sigma} \mathcal{F}_{\mu \nu} \mathcal{F}_{\rho \sigma}   \right) \,,  
 \end{split}
\end{equation}
in agreement with the form of  the local covariant action for  the trace anomaly~\cite{Giannotti:2008cv,Eling:2013bj}
\begin{equation}
\mathcal{W}_\mathrm{anom} = c \int d^4 x  \sqrt{-G}
\ln \frac{\bar{M}}{T}  
\, \mathcal{F}_{\mu \nu} \mathcal{F}^{\mu \nu} \,,  \qquad  c= -\frac{1}{48 \pi^2} \,. 
\end{equation} 
The trace of the stress tensor for chiral fermions  is given by  
\begin{equation}
G_{\mu \nu} \langle T^{\mu \nu}  \rangle  = -\frac{1}{48 \pi^2}  \mathcal{F}_{\mu \nu} \mathcal{F}^{\mu \nu} \,. 
\end{equation}

The results presented above correspond to a free theory of one left Weyl fermion. The functions $M_i(T,\nu)$ obtained with one free Dirac fermion are twice the expressions~(\ref{eq:M1})-(\ref{eq:M7}).

\section{Non-dissipative constitutive relations from the partition function}
\label{sec:constitutive_relations}

In this section, we use the partition function of eq.~(\ref{eq:W}) to
determine partially the non-dissipative part of the second order
constitutive relations in terms of the functions $M_i(T, \nu)$.  The
stress tensor and charge current of the fluid may be written in the form
\begin{equation}
\begin{split}
T^{\mu \nu} &= (\varepsilon + P) u^\mu u^\nu +  P G^{\mu \nu} + T_{(1)}^{\mu \nu}+  T_{(2)}^{\mu \nu}  + \ldots\, , \\
J^\mu &= \rho u^\mu + J_{(1)}^\mu + J_{(2)}^\mu + \dots \,, 
\end{split}
\end{equation}
where $\varepsilon$, $P$, $\rho$ and $u^\mu$ are the energy density, pressure, 
charge density and local fluid velocity respectively.
The subindex $(i)$ denotes the order in the derivative expansion.
While the first order constitutive relations have been extensively
considered in connection with the partition function, less attention
has been paid to the study of the second order terms, at least in the
case of a charged fluid.  Here, we will restrict to the parity even
terms that solely contain second order derivatives, i.e., terms of
$I_2$ type in the notation of refs.~\cite{Bhattacharyya:2012nq,
  Bhattacharyya:2014bha}.  In the linearized theory of hydrodynamic
fluctuations about the equilibrium these terms, together their parity
odd counterparts, are the most important.

In general, the determination of  non-dissipative parts in the constitutive relations  at a given order  
can be made by the comparison of the corresponding value of
$T^{\mu\nu}$ or $J^\mu$ evaluated at equilibrium with that obtained
from the partition function. The outline of the procedure may be sketched by   
\begin{equation}
\langle \mathcal{O}_i \rangle_\mathrm{eq} = 
   \delta (\mathcal{O}_\mathrm{perfect \,  fluid} +\mathcal{O}_{1} + \ldots + \mathcal{O}_{i-1} )   + \mathcal{O}_i \,, 
\end{equation}
where $\mathcal{O}_k$ corresponds to $T_{(k)}^{\mu \nu}$ or
$J_{(k)}^\mu$.  The left hand side is a specific variational
derivative of the partition function, and $\delta
(\mathcal{O}_\mathrm{perfect \, fluid} + \ldots)$ is a correction of
order $i$ due to all changes proportional to derivatives of the
background that must be evaluated in the constitutive relations of
lower orders.
In the Landau frame  we adopt, one also imposes the conditions 
\begin{equation}
T_{(i)}^{\mu \nu} u_\nu = 0, \qquad J_{(i)}^{\nu} u_\nu = 0 , \qquad i = 1, 2, \ldots . 
\end{equation}  
At the end, this procedure determines the transport coefficients, or
the susceptibilities, in $\mathcal{O}_i$ in terms of functions appearing
in the partition function.
At first order, one finds~\cite{Banerjee:2012iz}  
\begin{equation}
\begin{split}
T_{(1)}^{\mu \nu} &= 0 \,, \\ 
J_{(1)}^\mu &=  \xi_l l^\mu + \xi_B \mathcal{B}^\mu \,, 
\end{split}
 \end{equation}
where $l^\mu = \epsilon^{\mu \nu \rho \sigma} u_\nu \partial_\rho u_\sigma$,  
$\mathcal{B}^\mu =\tfrac{1}{2} \epsilon^{\mu \nu \rho \sigma} u_\nu \mathcal{F}_{\rho \sigma}$.
Here $\delta \mathcal{O}_\mathrm{perfect \,  fluid} $  receives a correction of 
 the fluid velocity $[\delta u_{(1)}]^i$, which at equilibrium is evaluated to a non-zero pseudo-vector, 
 while $\delta T_{(1)} = \delta \mu_{(1)} = 0$. 
 With the notation of~\cite{Bhattacharyya:2013ida}, 
the corresponding equations at second derivative order are 
\begin{align}
T_{00} \bigl|_\mathrm{eq} &= [u_{(0)}]_0^2 \, \delta\varepsilon_{(2)} + 2(\varepsilon + P) [u_{(0)}]_0 [\delta u_{(2)}]_0 \, , \\
T_{0}^{\;i} \bigl|_\mathrm{eq} &= (\varepsilon + P) [u_{(0)}]_0  [\delta u_{(2)}]^i \, , \\
T^{i j} \bigl|_\mathrm{eq} &=  \delta P_{(2)} g^{i j} + (\varepsilon + P) [\delta u_{(1)}]^i [\delta u_{(1)}]^j +
         T_{(2)}^{i j} \, , \\ 
J_{0} \bigl|_\mathrm{eq} &= [u_{(0)}]_0 \, \delta \rho_{(2)}  + \text{correction from $J_{(1)}$} \,, \\ 
J^{i} \bigl|_\mathrm{eq} &= \rho [\delta u_{(2)}]^i +  J_{(2)}^{i} +\text{correction from $J_{(1)}$}\, , 
\end{align} 
where 
\begin{equation}
\delta P_{(2)} = \frac{\partial P}{\partial\varepsilon} \delta\varepsilon_{(2)} + 
        \frac{\partial P}{\partial \rho} \delta \rho_{(2)} . 
\end{equation}
We have used some consequences of the  Landau frame condition evaluated at equilibrium, 
 which leads to  $J_{(2) \, 0} = T_{(2) \, 00} =  T_{(2)\,0}^{\quad \; i} = 0$, since $u_{(0)}^\mu =  e^{-\sigma}(1,0,0,0)$. 
 Note also that, since $[\delta u_{(1)}]^i$ is a pseudo-vector, the second order corrections
 that arise by substitution of  $[\delta u_{(1)}]^i$  in  the dissipative part of  $T_{(1)}^{\mu \nu}$ are parity odd. 
The implications of these contributions for  the parity odd transport coefficients have been recently studied in detail
 in ref.~\cite{Bhattacharyya:2013ida}.
With regard to the charged current, the parts termed as corrections from $J_{(1)}$ are parity even, 
but the explicit form,  that turns out to be quadratic in the anomaly coefficients, 
is not required for determining the linear terms in second derivatives. 

The most general non-dissipative form of the stress tensor and charge
current in the Landau frame at second order can be expressed as
\begin{equation}
\begin{split}
	\label{eq:constit}
T_{(2) \, \mu \nu} &= \Delta P \left(G_{\mu \nu} + u_\mu u_\nu \right) + 
T \left(\kappa_1 \tilde{R}_{\langle \mu \nu \rangle} + 
       \kappa_2 u^{\alpha} u^{\beta} \tilde{R}_{\langle \mu \alpha \nu\rangle \beta}  + 
       \kappa_3 \nabla_{\langle \mu} \nabla_{\nu \rangle} \nu \right)   \\ 
  & \quad + \text{combination of   six traceless bilinear tensors} \, , \\  
  J_{(2) \, \mu} &= \upsilon_1 P_{\mu \alpha}  u_\nu \tilde{R}^{\nu \alpha}  + 
   \upsilon_2 P_{\mu \alpha} \nabla_\nu \mathcal{F}^{\nu \alpha}  + 
   \text{combination of four bilinear vectors} \, ,
 \end{split}
\end{equation}
where $\nu(x) \equiv \mu(x)/T(x)$ reduces in equilibrium to
$A_0(\bm{x})/T_0$, being $\mu(x)$ the chemical potential.\footnote{The notation for the coefficients 
$\kappa_1$ and $\kappa_2$ is like that of ref.~\cite{Banerjee:2012iz}.} 
The
curvature quantities appearing in these constitutive relations are the
Ricci and Riemann tensors of the four-dimensional background. The
notation $X_{\langle \mu \nu \rangle}$ expresses the traceless and symmetric
combination transverse to $u^\mu$,
\begin{equation}
X_{\langle \mu \nu \rangle} \equiv P_{\mu}^\alpha P_{\nu}^\beta \left[ 
\frac{1}{2} (X_{\alpha \beta} + X_{\beta \alpha})  - 
\frac{1}{3} G_{\alpha \beta} P^{\gamma \theta}  X_{\gamma \theta} \right] \,, \qquad P_{\mu \nu} \equiv G_{\mu \nu} + u_\mu u_\nu \,.
\end{equation}
Although they do not play a role in our linear analysis, we also list  the non-dissipative bilinear tensor and vector 
quantities~\cite{Bhattacharyya:2014bha} appearing in eq.~(\ref{eq:constit}),  
\begin{equation}
\begin{split}
& 
\omega_{\langle \mu \alpha} \omega^{\alpha}_{\;\;\; \nu \rangle} \, , \quad
\omega_{\langle \mu \alpha} \mathcal{F}^{\alpha}_{\;\;\; \nu \rangle}  \, , \quad 
\mathcal{F}_{\langle \mu \alpha} \mathcal{F}^\alpha_{\;\;\; \nu \rangle} \, , \quad
 \nabla_{\langle \mu}  T \,  \nabla_{\nu \rangle}T \, , \quad
 \nabla_{\langle \mu}  T \,  \nabla_{\nu \rangle} \nu \, , \quad
 \nabla_{\langle \mu}  \nu \,  \nabla_{\nu \rangle} \nu \, , \\ 
& P_{\mu}^{\; \alpha} P^{\beta \nu}  \mathcal{F}_{\alpha \beta} \nabla_\nu T \, , \quad 
P_{\mu}^{\; \alpha} P^{\beta \nu}  \mathcal{F}_{\alpha \beta} \nabla_\nu \nu  \, , \quad
\omega_{\mu}^{\;\, \alpha} \nabla_\alpha T \, , \quad 
 \omega_{\mu}^{\;\, \alpha} \nabla_\alpha \nu \, ,
\end{split}
\end{equation}
where $\omega_{\mu \nu}$ is the vorticity tensor
\begin{equation}
\omega_{\mu \nu} \equiv \frac{1}{2} P_{\mu}^\alpha P_{\nu}^\beta \left( \nabla_\alpha u_\beta - \nabla_\beta u_\alpha \right) \, , 
\end{equation}
and $\mathcal{F}_{\mu \nu}$ is the gauge field strength. 
The correction to the pressure that includes second order derivatives with signature $+1$ under time reversal  is given by 
the combination
\begin{equation}
P_{(2)}=  \kappa_4 \tilde{R} + \kappa_5 D^2 T + \kappa_6 D^2 \nu \,  , 
\end{equation}
where $\tilde{R}$ and $D^2$ are the scalar curvature and the Laplacian
with respect to the four-dimensional metric $G^{\mu \nu}$,
respectively.~\footnote{The quantity $D^2 F$ becomes $\nabla^2 F + g^{i
    j} \partial_i \sigma \partial_j F$ for a time-independent scalar
  field, so $D^2 F\bigl|_\mathrm{eq} = \nabla^2 F + \text{bilinear
    terms in derivatives}$.}

The goal is to determine the coefficients $\kappa_i$ and $\lambda_j$
in eq.~(\ref{eq:constit}) by comparison with the partition function. By using the following variational derivatives
\begin{align}
T_{0 0} \bigl|_\mathrm{eq} &=  -\frac{T_0^2}{T \sqrt{g}}  \frac{\delta \mathcal{W}_2}{\delta \sigma} \notag \\ 
& =  T_0^2 \left(
-2 M_1 \nabla^2 T -  M_3 \nabla^2 \nu + \frac{\partial M_7^{\textrm{ren}}}{\partial T}  R \right) 
+ \text{bilinear terms in derivatives} \, ,  \\ 
T^{i j} \bigl|_\mathrm{eq} &=  -\frac{2 T}{\sqrt{g}} g^{i m} g^{j n} \frac{\delta \mathcal{W}_2}{\delta g^{m n}} \notag \\ 
& = 
-2 T M_7^{\textrm{ren}} \left( R^{i j} - \frac{g^{i j}}{2} R \right) + 2 T \frac{\partial M_7^{\textrm{ren}}}{\partial T} \left(\nabla^i \nabla^j  T -  
  g^{i j} \nabla^2 T \right)  \notag \\ 
& \quad + 2 T \frac{\partial M_7^{\textrm{ren}}}{\partial \nu} \left(\nabla^i \nabla^j  \nu -  
  g^{i j} \nabla^2 \nu \right) + \text{bilinear terms in derivatives} \,   , \\
T_0^{\;i} \bigl|_\mathrm{eq} &=    \frac{T}{\sqrt{g}} \left( 
      \frac{\delta \mathcal{W}_2}{\delta a_i}  -  A_0 \frac{\delta \mathcal{W}_2}{\delta A_i} \right)\notag \\ 
&= 2 T T_0  (2 \nu M_5 - M_6) \nabla_j F^{j i}  + 
      2 T  T_0^2  (-2 M_4^{\textrm{ren}} + \nu M_6) \nabla_j f^{j i} + \ldots \, , \\
J_{0} \bigl|_\mathrm{eq} &=  -\frac{T_0^2}{T \sqrt{g}}  \frac{\delta \mathcal{W}_2}{\delta A_0} \notag \\ 
& =  \frac{T_0}{T} \left(
 M_3 \nabla^2 T +2  M_2 \nabla^2 \nu - \frac{\partial M_7^{\textrm{ren}}}{\partial \nu}  R \right) + \ldots \, , \\ 
J^{i} \bigl|_\mathrm{eq} &=    \frac{T}{\sqrt{g}} \frac{\delta \mathcal{W}_2}{\delta A_i} \notag \\ 
&= -4 T M_5 \nabla_j F^{j i} - 2T  T_0 M_6 \nabla_j f^{j i} + \ldots \,  ,
\end{align}
together with the formulae 
\begin{align}
\tilde{R}_{\langle i j  \rangle} &= R_{i j}  - \frac{g_{i j}}{3} R + \frac{1}{T} \nabla_i \nabla_j T - \frac{g_{i j}}{3} \frac{\nabla^2 T}{T} + \ldots \, , \\ 
e^{-2 \sigma} \tilde{R}_{\langle i  0 j \rangle 0}  &= 
-\frac{1}{T} \nabla_i \nabla_j T + \frac{g_{i j}}{3} \frac{\nabla^2 T}{T} + \ldots \, ,  \\ 
\tilde{R} &= R + 2  \frac{\nabla^2 T}{T} + \ldots \, ,
\end{align}
we arrive at 
\begin{equation}
\label{eq:deltaP}
\begin{split}
\kappa_1 &= -2 M_7^{\textrm{ren}} \, , \\ 
\kappa_2 &= -2 M_7^{\textrm{ren}} - 2 T \frac{\partial M_7^{\textrm{ren}}}{\partial T} \, , \\ 
\kappa_3 &= 2 \frac{\partial M_7^{\textrm{ren}}}{\partial \nu} \,,  \\
P_{(2)}&= \left( \frac{T M_7^{\textrm{ren}}}{3} - 
           T^2 \frac{\partial M_7^{\textrm{ren}}}{\partial T} \frac{\partial P}{\partial \varepsilon}   - 
           \frac{\partial M_7^{\textrm{ren}}}{\partial \nu} \frac{\partial P}{\partial \rho}\right) \tilde{R} \\ 
            &\quad + \left(-\frac{4 T}{3}  \frac{\partial M_7^{\textrm{ren}}}{\partial \nu} + 
           T^2 M_3 \frac{\partial  P}{\partial \varepsilon}   +
           2 M_2 \frac{\partial P}{\partial \rho}\right) D^2  \nu \\ 
           &\quad + \left[ -\frac{2}{3} M_7^{\textrm{ren}} - \frac{4 T}{3} \frac{\partial M_7^{\textrm{ren}}}{\partial T}  + 
           2 \left( T^2 M_1 + T \frac{\partial M_7^{\textrm{ren}}}{\partial T} \right) \frac{\partial P}{\partial \varepsilon}   \right. \\ 
           &\quad + \left. \left(\frac{2}{T} \frac{\partial M_7^{\textrm{ren}}}{\partial \nu}  + 
           M_3 \right)\frac{\partial P}{\partial \rho}  \right] D^2 T \, + \ldots \,.
\end{split}
\end{equation}
Finally, by using the correction 
\begin{equation}
[\delta u_{(2)}]^i = -\frac{1}{e^{\sigma} (\varepsilon + P) } T_0^{\;i} \bigl|_\mathrm{eq} \, ,
\end{equation}
and the comparison of the vectors in the charged current  of eq.~(\ref{eq:constit}) with $J^{i} \bigl|_\mathrm{eq}$, we obtain 
\begin{equation}
\label{eq:lambda}
\begin{split}
\upsilon_1 &= 4 T^2 \left(2 \nu M_5 - M_6 \right) - \frac{8\rho}{\varepsilon + P} T^3 
 \left(M_4^{\textrm{ren}} + \nu^2 M_5   - \nu  M_6  \right) \,, \\
\upsilon_2 &=  - 4 T M_5  + \frac{2\rho}{\varepsilon + P} T^2 \left(2 \nu M_5 - M_6 \right) \,. 
\end{split}
\end{equation}
Note that the transport coefficients do not depend on $T_0$, although
$T_0$ appears explicitly in the partition function~(\ref{eq:W}). 

Eqs.~(\ref{eq:deltaP}) and (\ref{eq:lambda}) are general, and may be
applied to the massless theory we have considered above. In this case
we have $\partial P/\partial \varepsilon =1/3$ and $\partial
P/\partial \rho =0$, so that the results in
eqs.~(\ref{eq:M1})-(\ref{eq:M7}) produce $P_{(2)}=0$ to
linear order.  The effect of the trace anomaly appears in $P_{(2)}$ from
\begin{equation}
  \label{eq:trazeq}
 T^{i j} \bigl|_\mathrm{eq} g_{i j}  - T_{00} \bigl|_\mathrm{eq}  e^{-2 \sigma} =  
 T_{(2)}^{i j} g_{i j} = 3 P_{(2)} = - \frac{1}{48 \pi^2}  \mathcal{F}_{\alpha \beta} \mathcal{F}^{\alpha \beta}  \, ,
 \end{equation} 
 since the normalization condition $u^\mu u_\mu=-1$ implies that in equilibrium $[\delta u_{(1)}]_0 = 0$, and 
 \begin{equation}
 2 u_{(0)}^0 [\delta u_{(2)}]_0 + g_{i j}[\delta u_{(1)}]^i [\delta u_{(1)}]^j = 0 \,.
 \end{equation}

For the sake of completeness we show the explicit result of these transport coefficients obtained with the free field theory of Weyl fermions
\begin{equation}
\begin{split}
\kappa_1 &= \frac{T}{144} + \frac{1}{48\pi^2}\frac{\mu^2}{T} \,, \\
\kappa_2 &= 2 \kappa_1  \,, \\
\kappa_3 &=  -\frac{\mu}{24\pi^2}  \,, \\
\upsilon_1 &=  \frac{1}{2}\left(C+\frac{1}{3\pi^2} \right)\mu + \frac{\rho}{\varepsilon+P} \left[ -\frac{1}{2}\left(C+\frac{1}{6\pi^2}\right)\mu^2 + \left(C_2 - \frac{1}{36} \right)T^2  \right] \,, \\
\upsilon_2 &=  \frac{1}{24\pi^2} \left( \ln \frac{\bar{M}^2}{T^2} + Q\left(\frac{\mu}{T}\right) \right)  + \frac{\rho}{\varepsilon + P} \frac{1}{4}\left(C + \frac{1}{3\pi^2} \right)\mu \, , \label{eq:coef}
\end{split}
\end{equation}
where we have used that $\nu = \mu/T$.  
The only coefficient affecting second order derivatives  which
shows sensitivity to the renormalization scale is 
$\upsilon_2$ through their dependence on $-4 T M_5$.  
It is remarkable the absence of logarithms in
$\upsilon_1$ and $\kappa_{1,2,3}$, which in the former case is a
consequence of the particular combination of the $M$'s.  We also note
the presence of $C$ and $C_2$ in these second order results.

The form of other second order non-dissipative coefficients at weak coupling at zero chemical potential,
such as $\lambda_3$ and $\lambda_4$, 
\begin{equation}
T_{(2) \, \mu \nu} =   T \left( \lambda_3 \,  \omega_{\langle \mu \alpha} \omega^{\alpha}_{\;\;\; \nu \rangle}  + 
\lambda_4 \, \mathfrak{a}_{\langle \mu} \mathfrak{a}_{\nu \rangle} \right) \,, 
\qquad \mathfrak{a}_{\mu} = u^\alpha \nabla_\alpha u_\mu \,, 
\end{equation}
may be inferred  from the results derived in section 5 of ref.~\cite{Banerjee:2012iz}. 
There,  the parametrization of the partition function is made through three functions $\tilde{P}_i(T)$, 
whose relation to the $M_j$  is 
\begin{equation}
\begin{split}
\tilde{P}_1(T) &= -2 M_7(T, \nu=0), \\ 
\tilde{P}_2(T) &= -2 M_4(T, \nu=0), \\
\tilde{P}_3(T) &= -2 T^2 M_1(T, \nu =0) , 
\end{split}
\end{equation}
so, by using their formulae (5.8) and (5.15), one finds  
\begin{equation}
\begin{split}
\kappa_1 &=-2 M_7 , \\ 
\kappa_2 &=-2 M_7-2 T M_7'(T) , \\
\lambda_3 & = 16 T^2 M_4 - 6 M_7 - 2 T M_7'(T) , \\
\lambda_4 &= -2 T^2 M_1 + 4 T M_7'(T) + 2 T^2 M_7''(T) \, . 
\end{split}
\end{equation} 
With the values at hand for the $M_j$, it turns out that $\lambda_3= \lambda_4 = 0$. 

Let us compare these transport coefficients with some existing results
in the literature.  On the one hand $\kappa_1$, $\kappa_2$ and
$\lambda_3$ have been explicitly computed in ref.~\cite{Moore:2012tc}
in the case of a conformal fluid at zero chemical potential.  Our
values for $\kappa_1$ and $\kappa_2$ after setting $\mu=0$ agree with
the results in this reference.  The constraint $\kappa_2 = 2\kappa_1$
is also found in ref.~\cite{Banerjee:2012iz}.  Regarding $\lambda_3$,
as mentioned above we get a vanishing value, and this is in contrast
with the result obtained in ref.~\cite{Moore:2012tc}, where they find
$\lambda_3^{\textrm{Moore,Sohrabi}} = -T^2/24$ for a Weyl fermion, and
$-T^2/12$ for a Dirac fermion.  The difference is related to the
contribution proportional to~$ \text{rot}\, \bm{a} \cdot \langle
\tilde{\bm{J}}\rangle_1 $ in eq.~(\ref{eq:T00}), which seems to be not
included in the diagramatic computation of these authors.  In this
reference, triangle diagrams with cubic vertices in the fermion sector
are only computed, while that contribution is tied to a three-point
function from a seagull diagram with a quartic vertex.  All the
dependence on $C$ and $C_2$ in the second order coefficients comes
from this term, and so we would expect that after removing these
coefficients from the formulas, our result for $\lambda_3$ now agrees
with the result in \cite{Moore:2012tc}.  In fact after doing that,
they agree modulo a numerical factor, $\lambda_3|_{\mu=0, C=C_2 = 0} =
-2 \lambda_3^{\textrm{Moore,Sohrabi}}$.  This factor can only come
from the coefficient multiplying the term $\propto 1/T$ in
$M_4^{\textrm{ren}}$, but after a careful check we have not detected
any mistake in the computation.

Regarding the two terms involving the gauge field in eq.~(\ref{eq:constit}), an explicit computation of $\kappa_3$ has been performed in refs.~\cite{Erdmenger:2008rm,Banerjee:2008th,Megias:2013joa}, and $\upsilon_2$ in ref.~\cite{Megias:2013joa}, in the context of a holographic model in 5 dimensions with pure gauge and mixed gauge-gravitational Chern Simons terms. Taking care of the different notation used in these references, one can make the identification~$T \kappa_3 = \Lambda_5$ and $\upsilon_2 = -\xi_{10}$, where $\Lambda_5$ and $\xi_{10}$ are given by eqs.~(7.26) and (4.38) of ref.~\cite{Megias:2013joa} respectively. 
These coefficients receive contributions not induced by chiral anomalies, and so we cannot expect that the free field theory result of the present work agree with a strong coupling computation.
However, it is tempting to study the parametric dependence in $\mu$ and $T$ of these coefficients. 
Using the results above for $\kappa_3$ and $\upsilon_2$, and the explicit expressions of refs.~\cite{Erdmenger:2008rm,Banerjee:2008th,Megias:2013joa} for the analogous coefficients at strong coupling, one gets in the regime $\mu \ll  T$
\begin{align}
T \kappa_3 = - \frac{\mu T}{24\pi^2} \propto \Lambda_5  \,, \qquad \upsilon_2 \sim c(T) + \frac{5}{112\pi^4}\frac{\mu^2}{T^2}  \propto - \xi_{10} \,,
\end{align}
where in the free fermion computation $c(T)$ has a logarithmic dependence on $T$, 
while $c(T)$  is a constant in the holographic result, as the model of ref.~\cite{Megias:2013joa} 
doesn't include conformal symmetry breaking effects. So, apart from these considerations, we can confirm agreement in the parametric dependence between both approaches.


\section{Conclusion}
\label{sec:Conclusions}

In this paper we have addressed the computation of the thermal
partition function of an ideal gas of massless fermions on an
arbitrary stationary background in $3+1$ dimensions.  Using a
derivative expansion of the background fields, we have computed the
equilibrium values of the charged $U(1)$ current and energy-momentum
tensor.  We confirm the results previously reported in the literature
for the parity odd transport coefficients at first order, and find as
new results the parity even contributions at second order.  From this,
we derived the equilibrium partition function at second derivative
order, and showed that the renormalization effects of the conformal
anomaly mix with the chiral anomaly in some terms of the partition
function. However this mixture does not appear in the constitutive
relations. We have made the computation by using Pauli-Villars
regularization. It is remarkable that the finite contributions
from the regulators are crucial in order to obtain a consistent result for the partition function. 

The equilibrium partition function can only account for
non-dissipative effects, i.e.  it makes contact with transport
coefficients multiplying quantities that survive in equilibrium.
While first-order non-dissipative coefficients, like the chiral
magnetic and vortical conductivities, are ${\cal T}$-even and ${\cal
  P}$-odd, the situation at second order is however slightly
different.  Without violation of ${\cal T}$ invariance, the parity
violating part of the partition function at second order vanishes, so
the non-dissipative coefficients related to it are ${\cal T}$-even.
We examined the constitutive relations in the Landau frame, and we
derived the parametric dependence with temperature and chemical
potential of five transport coefficients: $\kappa_{1,2,3}$ and
$\upsilon_{1,2}$.  $\kappa_1$ and $\kappa_2$ are consistent with a
constraint previously reported in the literature, and the parametric
dependence in temperature and chemical potential of $\kappa_3$ and
$\upsilon_2$ agree with explicit results of these coefficients at
strong coupling. We have evaluated also two additional coefficients at
zero chemical potential, $\lambda_{3,4}$, and the result is that
$\lambda_4$~is vanishing as required by conformal invariance, and a
cancellation produces a zero value for $\lambda_3$.

\begin{acknowledgments}
We would like to thank Juan L. Ma\~nes for discussions, collaboration
on related topics, and for carefully reading the manuscript.  This
work has been supported by Plan Nacional de Altas Energ\'{\i}as
(FPA2011-25948 and FPA2012-34456), the Basque Government (IT559-10),
Spanish MICINN Consolider-Ingenio 2010 Program CPAN (CSD2007-00042)
and Centro de Excelencia Severo Ochoa Programme grant
SEV-2012-0234. E.M. would like to thank the Universidad del Pa\'{\i}s
Vasco UPV/EHU, Spain, for their hospitality and support during the
completion of parts of this work. The research of E.M. is supported by
the Juan de la Cierva Program of the Spanish MINECO.
\end{acknowledgments}


\appendix

\section{Free theory of Dirac fermions}
\label{sec:Dirac_fermions}

We show in this appendix some technical details of the free theory of Dirac fermions that are used in sections~\ref{sec:current_emt} and \ref{sec:green_function}. The action of the theory is
\begin{equation}
S = \int d^4x \sqrt{-G} \cL \,, \qquad \textrm{where} \qquad \cL = -i\bar\Psi\underline\gamma^\mu\nabla_\mu\Psi +im\bar\Psi\Psi \,,
\end{equation}
where $\bar\Psi = \Psi^\dagger \gamma^0$. The space-time dependent
Dirac matrices satisfy $\{\underline\gamma^\mu(x),
\underline\gamma^\nu(x)\}=2G^{\mu\nu}(x)$, and they are related to the
Minkowski matrices by $\underline\gamma^\mu(x)=e^\mu_a(x) \gamma^a$,
where $e_a^\mu(x)$ is the vierbein, $\{ \gamma^a, \gamma^b\} =
2\eta^{ab}$ and $\eta^{a b} = \text{diag}(-1,1,1,1)$. We choose the
Minkowski matrices in the Weyl representation
\begin{equation}
\gamma^0 = \left( \begin{array}{cc}
\mathbb{0}_{2\times2} &  \mathbb{1}_{2\times2} \\
-\mathbb{1}_{2\times2} &  \mathbb{0}_{2\times2} \end{array} \right)  \,, \qquad\quad 
\gamma^i =  \left( \begin{array}{cc}
\mathbb{0}_{2\times2} &  \sigma_i \\
\sigma_i    &  \mathbb{0}_{2\times2} \end{array} \right)  \,,\quad  i = 1,2,3 \,,
\end{equation}
where $\sigma_i$ are the Pauli matrices. The Dirac fields can be decomposed into left and right handed components, so that
\begin{equation}
\Psi = \left( \begin{array}{c} 
\psi_L \\
\psi_R  \\
\end{array} \right) \,, 
\qquad   \textrm{where} \qquad
\psi_{L (R)} = \left( \begin{array}{c}
\psi_1 \\
\psi_2
\end{array}
\right)_{L (R)}  \,. \label{eq:Psi}
\end{equation}
$\psi_{L (R)}$ are left (right) Weyl fermions of two components. The covariant derivative of the Dirac field is given by
\begin{equation}
\nabla_\mu \Psi = \left(\partial_\mu + \frac{1}{4} \omega_\mu^{\;\; a  b} \gamma_{a b} - i \A_\mu \right) \Psi \,, \qquad \gamma_{a b} = \frac{1}{2} [\gamma_a, \gamma_b]\ ,
\end{equation}
where, in the absence of torsion, the spin connection is related to the vierbein $e_a^\nu$ 
 by
\begin{equation}
\omega_\mu^{\;\; a  b} = -e^{b \nu} (\partial_\mu e_\nu^{a} - \Gamma_{\mu \nu}^{\sigma} e_\sigma^a) \,,
\end{equation}
and $\Gamma_{\mu\nu}^\sigma$ are the Christoffel symbols. 
The $U(1)$ current and the energy-momentum tensor are defined respectively as
\begin{equation}
J_\mu = \frac{1}{\sqrt{-G}} \frac{\delta S}{\delta \A^\mu} \,, \qquad 
T^{\mu\nu} = \frac{e_a^{\nu}}{\sqrt{-G}} \frac{\delta S}{\delta e_{a \mu} } \, .
\end{equation}
These formulas yield the following expressions
\begin{align}
J^0_{L,R} &= - \bar\Psi\underline\gamma^0 \cP_{L,R} \Psi  \,, \qquad 
J^i_{L,R} = - \bar\Psi\underline\gamma^i \cP_{L,R} \Psi \,,  \\
T_{\mu \nu} &=  \frac{i}{4} \bar{\Psi} \left[ \underline\gamma_\mu \overrightarrow{\nabla}_\nu  - \overleftarrow{\nabla}_\nu \underline\gamma_\mu + ( \mu \leftrightarrow \nu) \right ] \Psi 
\end{align}
where
\begin{equation}
\bar{\Psi}\, \overleftarrow{\nabla}_\mu = \bar{\psi}\left( \overleftarrow\partial_\mu  - \frac{1}{4} \omega_\mu^{\;\; a  b} \gamma_{a b} + i \A_\mu \right) \,,
\end{equation}
and the projectors on left and right handed components are defined as $\cP_{L,R} = \frac{1}{2}\left( 1 \pm \gamma_5\right)$. Using this technology, the explicit expressions of the currents and energy-momentum tensor can be obtained, and they are presented in section~\ref{sec:current_emt}.

\section{Technical details on the computation of the thermal expectation values: Matsubara sums}
\label{sec:technical}

Similarly as we do at first order in section~\ref{sec:1st_order}, the formulas for $\langle J_0\rangle$ and $\langle T_{00}\rangle$ at second order become
\begin{align}
\langle J_0 \rangle_2 &= T_0\sum_n\left[-e^\sigma \mathrm{tr}\, \cG_2(\bm  x, \bm  x, \omega_n)\right] \,,  \label{eq:1evJ0} \\
\langle T_{00} \rangle_2 &= T_0 \sum_n \left[ e^\sigma (A_0 + i \omega_n)\,  \text{tr}\, \cG_2(\bm  x, \bm  x, \omega_n) - 
    \frac{1}{4} e^{3 \sigma}  \epsilon^{ijk} \partial_j a_k\,\text{tr}\,\left[\sigma_i\, \cG_1(\bm  x,\bm  x, \omega_n)\right]  \right]\,,  \label{eq:1evT00} 
\end{align}
where  $\omega_n = \frac{2\pi}{\beta}\left( n + \frac{1}{2}\right)$ are the fermionic Matsubara frequencies. 
The traces that will be relevant for this computation are
\begin{equation}
\begin{split}
\mathrm{tr} \, \cG_2(\bm x, \bm x, \omega_n ) &= \frac{e^{-2\sigma}}{96\pi^{3/2}} \int_0^\infty \frac{d\rho}{\sqrt{\rho}}  
 e^{b^2\rho}\bigg\{ 
-2 \nabla^i \sigma \nabla_i \sigma \bigl[ 3\tilde\omega_n + 2\rho e^{-2\sigma}\tilde\omega_n^3 \bigr]   \\
&\quad +  \nabla^i\sigma  \nabla_i A_0 \bigl[ 15 + 14 \rho e^{-2\sigma}\tilde\omega_n^2 \big] \\ 
&\quad - \nabla^i A_0 \nabla_i A_0 \rho e^{-2\sigma}\big[ 11\tilde\omega_n + 2\rho e^{-2\sigma}\tilde\omega_n^3 \bigr]  \\
&\quad  -\frac{1}{2}F_{ij}F^{ij} \rho \bigl[ 11 \tilde\omega_n + 2\rho^2 e^{-2\sigma} \tilde\omega_n^3 \bigr]   \\
&\quad + f_{ij} F^{ij} \bigl[-e^{2\sigma} - 11\rho A_0 \tilde\omega_n + 9 \rho \tilde\omega_n^2 - 2\rho^2e^{-2\sigma}A_0 \tilde\omega_n^3  + 2\rho^2 e^{-2\sigma} \tilde\omega_n^4 \bigr]   \\
&\quad +  f_{ij}f^{ij} \bigl[ -e^{2\sigma}A_0 + \frac{1}{16}\left( 13 e^{2\sigma} - 88\rho A_0^2 \right) \tilde\omega_n + 9\rho A_0 \tilde\omega_n^2 \\ 
&\quad - \frac{\rho}{8}\left(31 + 8\rho e^{-2\sigma}A_0^2 \right) \tilde\omega_n^3 
 + 2\rho^2 e^{-2\sigma} A_0 \tilde\omega_n^4 - \rho^2 e^{-2\sigma} \tilde\omega_n^5  \bigr]  \\
&\quad + \frac{1}{6} R \bigl[ 4\tilde\omega_n - 43\rho e^{-2\sigma}\tilde\omega_n^3 - 18\rho^2 e^{-4\sigma}\tilde\omega_n^5 \bigr]  \bigg\} \,, 
\end{split}
\end{equation}
where $\tilde\omega_n \equiv A_0 + i\omega_n $, in addition to $\mathrm{tr} \left[ \sigma_i \, \cG_1(\bm x, \bm x, \omega_n) \right]$ given by eq.~(\ref{eq:trsiG1}). In these formulas $R$ is the Ricci scalar from $g_{ij}$.

The summations over Matsubara frequencies are performed in the following way. We define the function
\begin{equation}
F(\rho,A_0) := T_0 \sum_n \, e^{b^2\rho} = \frac{e^{-m^2\rho + \sigma}}{2\sqrt{\pi\rho}} \vartheta_3\left( \frac{1}{2}\left( \pi - iA_0\beta \right) , e^{-\frac{e^{2\sigma}\beta^2}{4\rho}} \right) \,, \label{eq:F}
\end{equation}
where $b^2 = -m^2 + e^{-2\sigma} \tilde\omega_n^2$ and $\vartheta_3$ is a Jacobi $\Theta$ function, which admits the expansion
\begin{equation}
\vartheta_3(u,q) = 1 + 2\sum_{n=1}^\infty q^{n^2}\cos\left( 2nu\right) \,. \label{eq:theta3}
\end{equation}
Then the several powers in Matsubara frequencies 
\begin{equation}
F_m(\rho,A_0) := T_0 \sum_n \, e^{b^2\rho} \, \tilde\omega_n^m \,,
\end{equation}
can be obtained straightforwardly from appropriate combinations of derivatives of $F$. In particular, for the computation of the thermal expectation values of $J_0$ and $T_{00}$ up to second order in derivatives we need powers of $\tilde\omega_n$ up to order $m=6$. We find
\begin{align}
F_1(\rho,A_0) &= \frac{e^{2\sigma}}{2\rho} \frac{\partial F}{\partial A_0} \,, \label{eq:F1} \\
F_2(\rho,A_0) &= e^{2\sigma} \left( \frac{\partial F}{\partial \rho} + m^2 F\right) \,, \\
F_3(\rho,A_0) &= \frac{e^{4\sigma}}{2\rho^2} \left(  \rho \frac{\partial^2 F}{\partial\rho\partial A_0}  - (1-m^2\rho) \frac{\partial F}{\partial A_0}  \right) \,, \\
F_4(\rho,A_0) &=  e^{4\sigma}  \left(  \frac{\partial^2 F}{\partial\rho^2} + 2m^2 \frac{\partial F}{\partial\rho} + m^4 F \right) \,, \\
F_5(\rho,A_0) &= \frac{e^{6\sigma}}{2\rho^3} \left( \rho^2 \frac{\partial^3 F}{\partial\rho^2\partial A_0} - 2\rho(1-m^2\rho) \frac{\partial^2 F}{\partial\rho\partial A_0} + (2-2m^2\rho + m^4\rho^2) \frac{\partial F}{\partial A_0} \right) \,, \\
F_6(\rho,A_0) &= \frac{e^{6\sigma}}{4\rho^2} \bigg( 4\rho^2\frac{\partial^3 F}{\partial\sigma^3} + 3e^{2\sigma} m^2 \frac{\partial^3 F}{\partial\rho \partial A_0^2} -30m^2\rho\frac{\partial F}{\partial \sigma} \nonumber \\
&\quad - 2 m^2(3 + 12m^2\rho -2m^4\rho^2 )F \bigg)  \,. \label{eq:F6}
\end{align}
The term with the summation $\sum_{n=1}^\infty$ in the rhs of eq.~(\ref{eq:theta3}) is responsible for the finite temperature and chemical potential contributions in the thermal expectation values, and they are never affected by UV divergences when integrating in the proper time $\rho$. The other term ``$1$'' leads to the vacuum contributions which are affected by these divergences. In view of eqs.~(\ref{eq:F}) and (\ref{eq:F1})-(\ref{eq:F6}) it is clear that the vacuum contributions can only appear in those terms with even powers of $\tilde\omega_n$, as these terms include contributions with no derivatives with respect to $A_0$. Finally both vacuum and finite temperature contributions are affected by IR divergences when integrating in $\rho$, as explained in section~\ref{sec:2nd_order}.

\bibliographystyle{JHEP}
\bibliography{refs}

\end{document}